%% file: cdf_isr.tex
\RequirePackage{fix-cm}
\documentclass[reprint,superscriptaddress,nofootinbib,showpacs,amsmath,amssymb,aps,prd,
]{revtex4-2}
\pdfoutput=1 			
\RequirePackage{graphicx}
\usepackage{graphicx}
\usepackage{dcolumn}
\usepackage{bm}
\usepackage{lineno}
\usepackage{setspace}
\usepackage{makecell}
\usepackage{multirow}
\usepackage{array}
\usepackage{amsmath}
\newcommand{\myhdashline}[1]{\multispan{#1}\unskip
  \vrule height \arrayrulewidth width 4pt \hskip 2pt
  \xleaders
  \hbox{\hskip 2pt
    \vrule height \arrayrulewidth width 4pt \hskip 2pt}
  \hfill \hskip 2pt
  \vrule height \arrayrulewidth width 4pt \cr
  \noalign{\vskip-\arrayrulewidth}
}
\newcommand{\RNum}[1]{\uppercase\expandafter{\romannumeral #1\relax}}
\newcommand{\qq}{\ifmmode{Q^{2}}
  \else
  \mbox{$Q^{2}$}
  \fi
}
\newcommand{\met}{
  {\not\!\!E_T}
}
\newcommand{\zg}{\ifmmode{Z/\gamma^*}	
  \else	
  \mbox{$Z/\gamma^*$}
  \fi	
}
\newcommand{\pt} {\ifmmode {p_{T}}
  \else 
  \mbox{$p_{T}$}
  \fi
}
\newcommand{\pte} {\ifmmode {\langle p_{T} \rangle}
  \else 
  \mbox{$\langle p_{T} \rangle$}
  \fi
}
\newcommand{\ptzg} {\ifmmode {p_{T}^{\mathrm{DY}}}
  \else 
  \mbox{$p_{T}^{\mathrm{DY}}$}
  \fi
}
\newcommand{\ptzge} {\ifmmode {\langle p_{T}^{\mathrm{DY}} \rangle }
  \else 
  \mbox{$\langle p_{T}^{\mathrm{DY}} \rangle$}
  \fi
}
\newcommand{\LL} {\ifmmode {\ell\ell}
  \else 
  \mbox{$\ell\ell$}
  \fi
}
\newcommand{\ptll} {\ifmmode {p_{T}^{\ell\ell}}
  \else 
  \mbox{$p_{T}^{\ell\ell}$}
  \fi
}
\newcommand{\ptlle} {\ifmmode {\langle p_{T}^{\ell\ell} \rangle}
  \else 
  \mbox{$\langle p_{T}^{\ell\ell} \rangle$}
  \fi
}
\newcommand{\mm} {\ifmmode {\mu\mu}
  \else 
  \mbox{$\mu\mu$}
  \fi
}
\newcommand{\ptee} {\ifmmode {p_{T}^{ee}}
  \else 
  \mbox{$p_{T}^{ee}$}
  \fi
}
\newcommand{\mzpole} {
  \ifmmode {m_{Z\mathrm{,pole}}}
  \else \mbox{$m_{Z\mathrm{,pole}}$}
  \fi
}
\newcommand{\mzpolesq} {\ifmmode {m^{2}_{Z\mathrm{,pole}}}
  \else 
  \mbox{$m^{2}_{Z\mathrm{,pole}}$}
  \fi
}
\newcommand{\mzg} {\ifmmode {m_{\mathrm{DY}}}
  \else 
  \mbox{$m_{\mathrm{DY}}$}
  \fi
}
\newcommand{\mzgsq} {\ifmmode {m_{\mathrm{DY}}^{2}}
  \else 
  \mbox{$m_{\mathrm{DY}}^{2}$}
  \fi
}
\newcommand{\mzgsqe} {\ifmmode {{\langle m_{\mathrm{DY}} \rangle}^{2}}
  \else 
  \mbox{${\langle m_{\mathrm{DY}} \rangle}^{2}$}
  \fi
}
\newcommand{\mzge} {\ifmmode {\langle m_{\mathrm{DY}} \rangle}
  \else 
  \mbox{$\langle m_{\mathrm{DY}} \rangle$}
  \fi
}

\newcommand{\mmm} {\ifmmode {m_{\mu\mu}}
  \else 
  \mbox{$m_{\mu\mu}$}
  \fi
}
\newcommand{\ptmm} {\ifmmode {p_{T}^{\mu\mu}}
  \else 
  \mbox{$p_{T}^{\mu\mu}$}
  \fi
}
\newcommand{\mee} {\ifmmode {m_{ee}}
  \else 
  \mbox{$m_{ee}$}
  \fi
}
\newcommand{\mll} {\ifmmode {m_{\LL}}
  \else 
  \mbox{$m_{\LL}$}
  \fi
}
\newcommand{\mlle} {\ifmmode {\langle \mll \rangle}
  \else 
  \mbox{$\langle \mll \rangle$}
  \fi
}
\newcommand{\et} {\ifmmode {E_{T}} 
  \else  
  \mbox{$E_{T}$}
  \fi
}
\newcommand{\alps}{\ifmmode {\alpha_{S}}
  \else    \mbox{$\alpha_{S}$}
  \fi
}
\newcommand{\ifb}{\ifmmode {\mathrm{fb^{-1}}}
  \else    \mbox{$\mathrm{fb^{-1}}$}
  \fi
}
\newcommand{\msq}{\ifmmode {m_{\LL}^{2}}
  \else \mbox{$m_{\LL}^{2}$}
  \fi
}
\newcommand{\gevc}{\ifmmode {\rm{GeV}/\it{c}}
  \else \mbox{GeV/$\it{c}$}
  \fi
}
\newcommand{\gevcc}{\ifmmode {{\rm GeV}/c^2}
  \else \mbox{GeV/$c^2$}
  \fi
}

\begin{document}

\title{A novel measurement of initial-state gluon radiation in hadron collisions using Drell-Yan events
}


\input{author_prd}



\date{Received: date / Accepted: date}

\begin{abstract}
A study of initial-state gluon radiation (ISR) in hadron collisions is presented using Drell-Yan (DY) events produced in proton-antiproton collisions by the Tevatron collider at a center-of-mass energy of 1.96~TeV.
This paper adopts a novel approach which uses the mean value of the \zg transverse momentum \ptzge in DY events as a powerful observable to characterize the effect of ISR.
In a data sample corresponding to an integrated luminosity of 9.4~$\ifb$ collected with the CDF \RNum{2} detector, $\ptzge$ is measured as a function of the \zg invariant mass.
It is found that these two observables have a dependence,
$\ptzge = -8 + 2.2 \ln{\mzgsq}~[\gevc]$, where \mzg is the value of the \zg mass measured in units of $\gevcc$.
This linear dependence is observed for the first time in this analysis. It may be exploited to model the effect of ISR and constrain its impact in other processes.

\end{abstract}

\maketitle

\section{Introduction}
\label{sec:intro}
Since the discovery of the Higgs boson~\cite{Higgs1,Higgs2} in elementary particle physics, the search for physics beyond the Standard Model (SM) has become the main focus of attention.
To identify small deviations from the SM expectations due to beyond the SM (BSM) physics, a precise understanding of the SM processes is required.
At hadron colliders SM interactions are commonly accompanied by clusters of final-state hadrons (jets) generated from initial-state gluon radiation (ISR). A large fraction of these hadrons have low transverse momemtum ($\pt$) and are difficult to simulate and measure correctly.
An accurate modeling of ISR is essential in BSM searches at the LHC, since many relevant models of BSM physics predict the production of undetectable particles (such as those expected from dark-matter candidates) whose presence can be inferred by triggering on single isolated jets produced by ISR. A detailed understanding of ISR effects would also benefit a precise measurement of the top-quark mass by enabling accurate modeling of the top-quark transverse momentum distribution.

In hadron collisions, quantum-chromodynamics (QCD) gluon emissions from the interacting partons are conventionally classified into two categories: hard and soft/collinear emissions.
The hard QCD emissions are approximately described by perturbative QCD~\cite{zjet}.
The soft/collinear QCD emissions are mostly non-perturbative in nature and they are absorbed into the parton distribution functions (PDFs).
Parton-shower algorithms have been developed to approximate the physics of the non-perturbative emissions using the DGLAP equations~\cite{DGLAP1,DGLAP2,DGLAP3,DGLAP4}. The algorithms simulate the general features of non-perturbative QCD, but their accuracy may be insufficient for high-precision analyses~\cite{topmass} or some BSM particle searches.
A systematic approach to study non-perturbative emissions is required.

The PDFs and QCD radiation are closely associated. The DGLAP equations describe the evolution of PDFs, 
which have been studied extensively in lepton-nucleon inelastic-scattering experiments~\cite{hera}.
The DGLAP equations describe the change in the quark density function of incoming quarks due to QCD processes:
\begin{linenomath}
\begin{equation}
 \frac{dq(x,Q^2)} {d \ln{Q^2}} = \int_{x}^{1} \frac{dy}{y} \alpha _s(Q^2)  P_{q \rightarrow qg}\left(\frac{x}{y},Q^2\right) q(y,Q^2) .
\end{equation}
\end{linenomath}
Here, $\alpha _s$ is the strong coupling constant which is a function of the energy scale, $Q^2$. 
The term $q(y,Q^2)$ is the quark density function with momentum fraction $y~(> x)$ in the proton, and $P_{q \rightarrow qg}$ is the QCD splitting function that gives the probability
for the incoming quark to split into a quark and a gluon. Since all of these QCD processes are described by the DGLAP equations which have a logarithmic $Q^2$ dependence, this study investigates the effects of QCD ISR as a function of $Q^2$.

The Drell-Yan (DY) production of lepton pairs $(p \bar{p} \to \zg \to  ee, \mm)$ is ideal to study QCD ISR. At the Born level, DY lepton pairs are produced with zero transverse momentum. The emission of one or more gluons from the initial-state quarks gives rise to a transverse momentum for these quarks which then creates a non-zero transverse momentum for the DY lepton pair. Thus, a transverse momomentum of the DY lepton pair is a good observable to study the effect of QCD ISR. The final state of the DY lepton pair is free from final-state QCD radiation. This simplifies the interpretation of the measurements. In addition, the parton-parton energy scale $Q^2$ is characterized by the squared dilepton mass $m_{\LL}^2$.

\begin{figure}[b]
  \centering
  \includegraphics[width=0.5\textwidth]{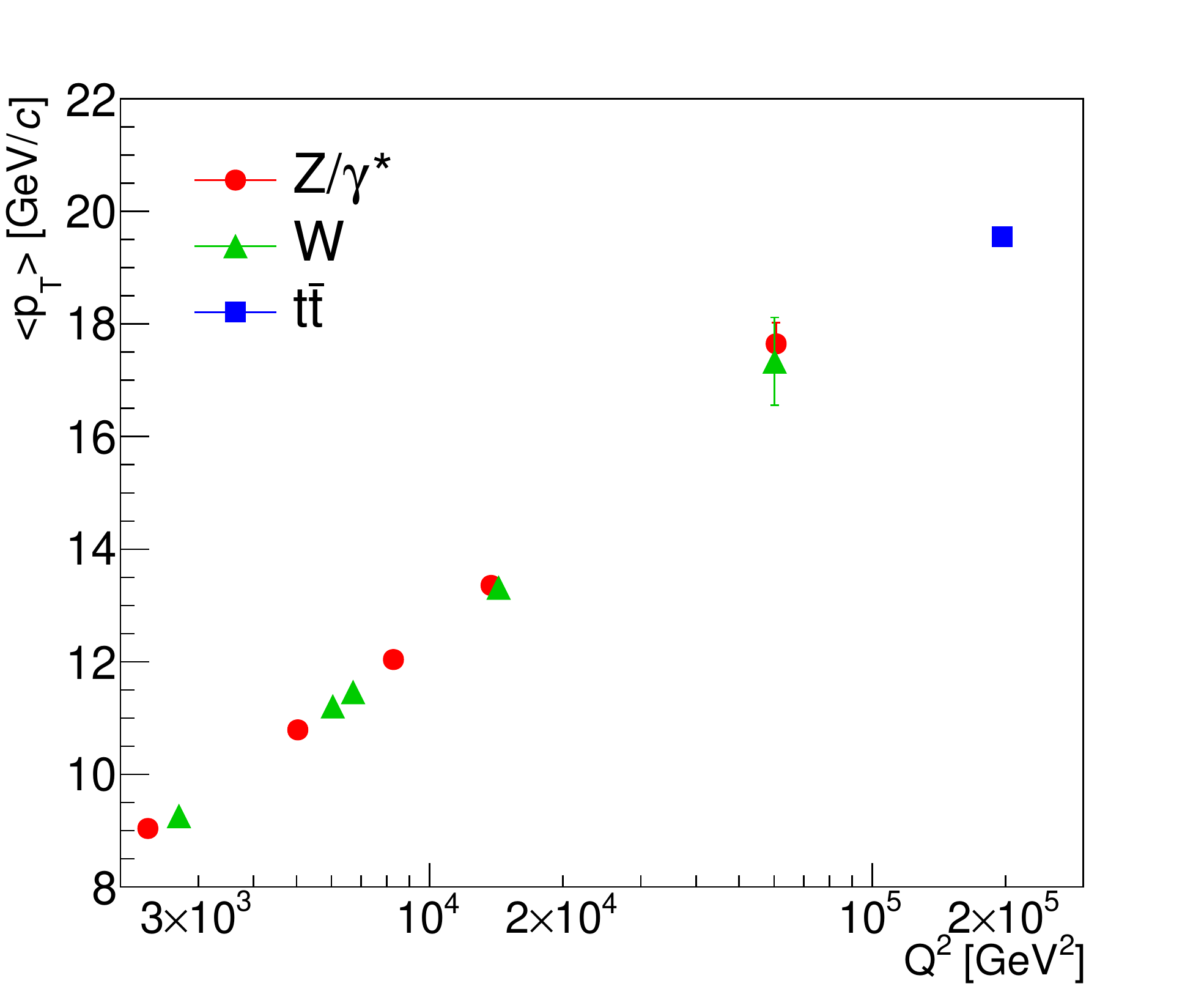}
  \caption{Predicted $\pte$ of the hard process system ($\zg$, $W$ or $t\bar{t}$) as a function of the energy scale $Q^2$. Here, $\pte$ is the truncated mean of the $\pt$ distribution with $\pt < 100~\gevc$ and $Q^2$ is set as the squared mass of $\zg$, $W$ or $t\bar{t}$, respectively. The values shown are obtained from {\sc Pythia8} simulations of inclusive hadroproduction of $Z/\gamma^*$, $W$ bosons, and top-antitop quark pairs.
  }
  \label{plot_intro_otherprocess}
\end{figure}

Using {\sc Pythia8}~\cite{py8} simulations, it is found that the truncated mean of $\pt$ distributions $\pte$ for the DY process and other DY-type process (such as $W$ boson or top-quark pair production) can be described with a function linear in the logarithm of the energy scale of the hard process with universal slope, as a consequence of the DGLAP evolution. This universality is illustrated in Fig.~\ref{plot_intro_otherprocess}, which shows the prediction for $\pte$ of different processes, exhibiting a common dependence on the energy scale $Q^2$. Since the recoil of the hard process system due to ISR is the main factor of $\pte$, it can be used as a powerful observable to probe the effect of ISR.

This paper presents a novel approach to characterize the effect of ISR, which can be observed in the average \pt of DY lepton pairs as a function of the energy scale $Q^2$. The measurements of the average \pt at various energy scales are performed using proton-antiproton ($p\bar{p}$) collision data at center-of-mass energy $\sqrt{s} =$ 1.96 TeV produced by the Tevatron collider and collected with the CDF \RNum{2} detector.
The measured values are corrected for QED final-state radiation (FSR) effects and the experimental effects such as detector acceptance and resolution.
Using such measurements a relationship between the \pt and mass-scale observables in the DY process can be established. This may improve modeling of ISR phenomenology as well as constrain its impact on the measurement of other processes in hadron colliders.

\section{The Experimental Apparatus}
\label{sec:experiment}

The CDF \RNum{2} detector is a solenoidal magnetic spectrometer surrounded by calorimeters and muon detectors, which operated at the Tevatron proton-antiproton collider from 2001 until 2011. CDF \RNum{2} uses a cylindrical coordinate system with the positive $z$-axis along the proton beam direction. For particle trajectories, the polar angle $\theta$ is relative to the proton direction and the azimuthal angle $\phi$ is oriented about the beamline axis with $\pi/2$ being vertically upwards.
Detector coordinates are specified as ($\eta, \phi$), where $\eta$ is the pseudorapidity defined as $-\ln \tan (\theta/2)$. The detector is described in detail in Ref.~\cite{cdfdyxsec}.

The beam pipe is surrounded by a 2 m long silicon vertex-tracker covering a pseudorapidity range of $|\eta| < 2$~\cite{svx}. The central charged-particle tracking detector, which is a 3.1 m long open-cell drift chamber, extends radially from 0.4 to 1.4 m covering the range $|\eta| < 1.0$~\cite{cot}.
Both trackers are positioned in a 1.4 T axial magnetic field produced by a superconducting solenoid surrounding the outer radius of the drift chamber. Outside the solenoid is a central barrel calorimeter in the region $|\eta| < 1.1$~\cite{cem,cha/wha}. The forward end-cap regions are covered by plug calorimeters in the regions $1.1 < |\eta| < 3.5$~\cite{pem}. Muon detectors are the outermost charged-particle trackers and cover the region $|\eta| < 1.5$~\cite{muonupgrade}.

The data were collected using a three-level electronics system (trigger).
The first level, relying on special-purpose processors, and the second level,
using a mixture of dedicated processors and fast software algorithms, reduce the event accept-rate to
a level manageable by the data acquisition system. The accepted events are processed online
at the third-level trigger~\cite{l3} with fast reconstruction algorithms~\cite{ev},
and are recorded for offline analysis.

\section{Event Selection}
\label{sec:event}

The DY candidate muon-pairs were accepted online by a single-muon trigger~\cite{cdfdyxsec} with track $\pt$ threshold of 18 GeV/$c$. Electron candidate~\cite{cdfdyxsec} pairs were accepted online by single- and double-electron triggers~\cite{cdfdyxsec,dyee}.
The single-electron trigger requires at least one electron candidate with transverse energy $\et$ greater than 18 GeV and an associated track with \pt larger than 9~$\gevc$. For the double-electron trigger, events are accepted when containing at least two electron candidates without requiring an associated track.

Events collected are further required to pass the following offline-selection criteria. Muon candidates are required to have matching track elements in the muon chambers (except when the track extrapolates outward to an uninstrumented region (gap) of the muon detector),
and an energy deposition in the calorimeters consistent with that for a minimum-ionizing particle~\cite{cdfdyxsec}.
Track-quality selections are also imposed on the candidates.
The selection of muon pairs requires two oppositely-charged muon candidates with
$|\eta| <$ 1.5. The muon candidate with leading (sub-leading) \pt is required to be larger than 20~(12)~$\gevc$.

Electron candidates are either reconstructed in the central electromagnetic calorimeter or in the forward region covered by the plug electromagnetic calorimeter~\cite{cdfdyxsec}.
Two levels of identification criteria are used for central-electron candidates. High-quality criteria select `tight' central electron (TCE) candidates by requiring a matching good-quality track in the tracking chamber and a shower profile consistent with that of an electron in the electromagnetic calorimeter. Less stringent criteria select 'loose' central electron (LCE) candidates by relaxing some track- and shower-quality requirements.
Plug-electron identification uses plug electromagnetic calorimeter information and requires an associated track reconstructed in the silicon tracker.

Electron pairs are classified into three topologies depending on where each electron candidate is reconstructed, central-central (CC, 40\% of selected candidates), central-plug (CP, 46\%), and plug-plug (PP, 14\%).  For CC-topology pairs, at least one electron candidate should pass TCE identification. The two electron candidates are required to be oppositely charged, and the candidate with leading (sub-leading) $\et$ is required to have  $\et >$ 25 (15) GeV. In the CP-topology, the central-electron candidate is required to pass TCE criteria. Both electron candidates are required to have $\et >$ 20~GeV.
The plug-electron pair candidates (PP-topology) are required to pass the plug-electron identification criteria and to have $\et >$ 25~GeV.
In addition, PP-topology pairs are requested to be contained in the same side plug detector to reduce the QCD background, because the QCD multijet events tend to have larger $\eta$ difference between the two leptons than DY events.
The plug-electron candidates have a poor charge identification because electron tracks in the forward region are constructed only using silicon detector information due to the small coverage of the drift chamber. Thus, the opposite-charge requirement is not applied to CP- and PP-topology pairs.
In order to reduce misreconstructed events and contributions from background processes,
the transverse energy imbalance ($\met$)~\cite{met} associated to the selected dielectron events is required to be lower than 40~GeV.

Dielectron events produced at the $Z$-mass peak may be reconstructed at much lower masses if the electron loses a significant portion of its energy by radiating photons.
It is difficult to simulate photons from the QED FSR process at small dielectron masses ($\mee$), where the effect is comparatively more important due to this migration.
To suppress the migration effects, the following additional selection criteria are applied to the events with dielectron invariant mass $\mee < 80~\gevcc$ based on a study using a DY simulated sample:
\begin{linenomath}
\[
\begin{array}{l}
  \begin{array}{l}
    \text{Reject\ if\ } |\Delta\phi(e_1,e_2)-\pi|<0.25 \rm{\ and \ } \\
    \Big\{
    \begin{array}{l}
      \Delta\pt(e_1,e_2)>15 \ \text{\gevc,\ when}\ \met < 15 \  \text{GeV}\\[1ex]
      \Delta\pt(e_1,e_2)>10 \ \text{\gevc,\ when}\ \met > 15 \ \text{GeV}.\\
    \end{array}
  \end{array}
\end{array}
\]
\end{linenomath}
Here, $e_1$ and $e_2$ are the leading and the sub-leading electrons in \pt, $\Delta\phi(e_1,e_2)$ is the azimuthal opening angle of electron pairs, and $\Delta\pt(e_1,e_2)$ is the scalar \pt difference of the two electrons.

Differences in acceptance and efficiency between simulation and data are
corrected by applying scale factors to the event yields reconstructed in simulation.
The trigger and identification scale factors for electrons are obtained using unbiased electron samples in data as functions of pseudorapidity, $\et$, number of extra vertices
in the event, and the time when the events were collected.
The scale factors for the muons depend on the detector topology of the muon pair and data taking time.
For the selected electron- and muon-pair events of the simulation, the number of
primary vertices and the distribution of the vertices along the beamline
are weighted to match their corresponding distributions in data. 

As this analysis focuses on QCD ISR mostly from non-perturbative QCD emissions, all candidates are required to have the dilepton system transverse momentum $\ptll < 100~\gevc$.
This selection also has the benefit of removing events with poorly reconstructed muons, which tend to have large measured $\ptll$ values.

\section{Drell-Yan signal events and backgrounds}
\label{sec:background}

\begin{figure}	
  \centering
  \includegraphics[width=0.5\textwidth]{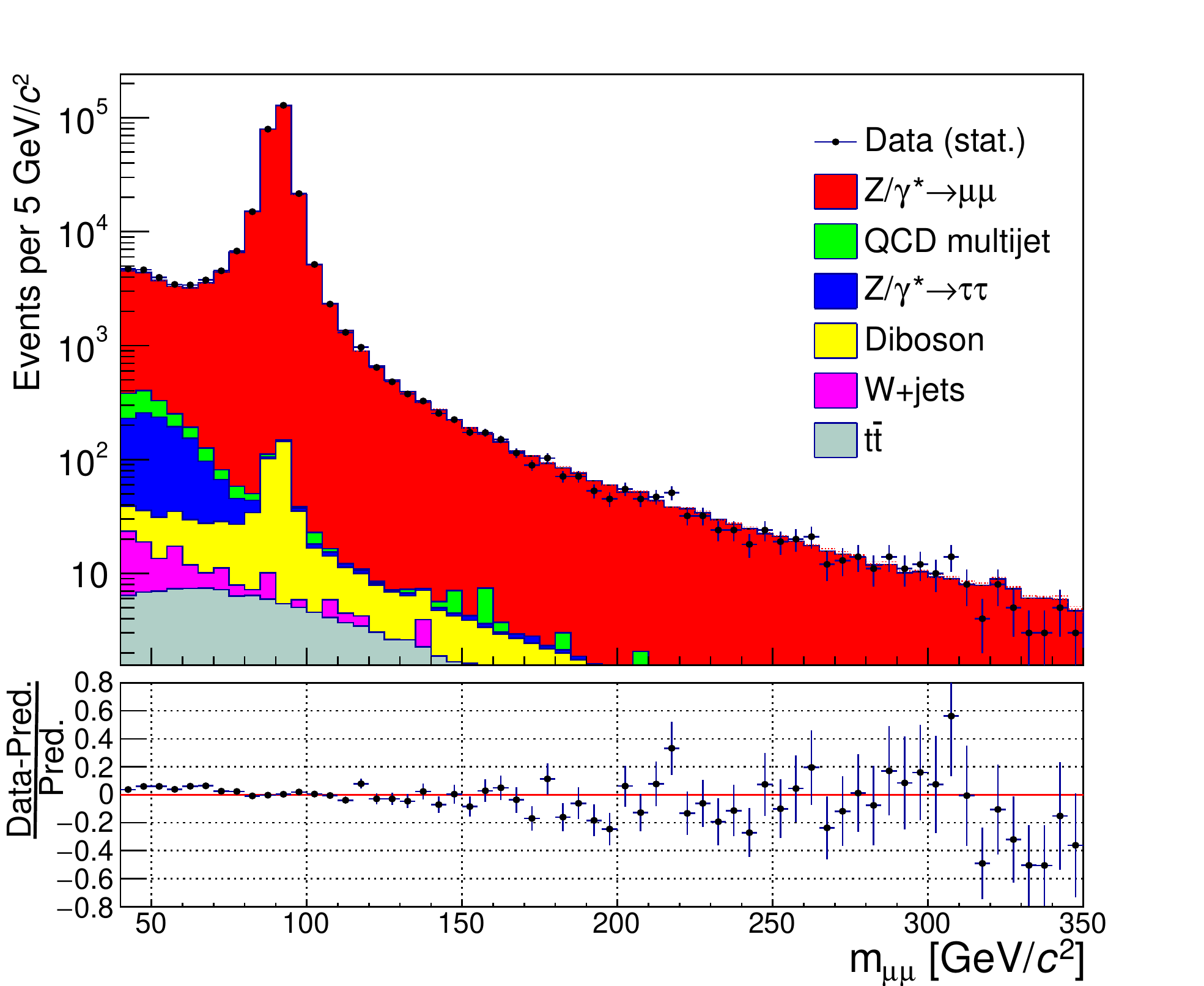}
  \caption{Observed dimuon mass distribution in data (circles) compared with the distributions from expected SM contributions (stacked). The normalized residuals of the data and the SM prediction are shown in the lower panel. Only statistical uncertainties are considered in both panels.}
  \label{figure:background_plot_muon}
\end{figure}

The selected dimuon events are dominated by the $\zg \to \mu\mu$ signal process, and there are also contributions from $\zg \to \tau\tau$, diboson ({\it WW}, {\it WZ}, {\it ZZ}), $t\bar{t}$, $W+$jets, and QCD multijet processes, which are considered as backgrounds.
They are all estimated using simulation with the exception of the QCD multijet process.
The simulated samples are generated using {\sc Pythia6}~\cite{pythia6} with tune AW~\cite{tuneaw}
and processed through the {\sc Geant3}-based~\cite{geant3} CDF \RNum{2} detector simulation. The total cross sections corresponding to the size of simulated samples are normalized using NLO QCD calculations respectively~\cite{dyee,vvxsec,ttxsec}. Following the procedure described in Ref.~\cite{dymm}, the multijet background for dimuon events is determined from the data using events with same-sign muon pairs assumed to have same kinematics with opposite-sign events.
Fig.~\ref{figure:background_plot_muon} shows the distributions of the dimuon mass, $\mmm$, for the SM processes considered and for the experimental data. 
The dimuon event yields for the data and the expected SM contributions are shown in Table~\ref{table:background_plot_muon} for various dimuon-mass ranges.
Given that only statistical uncertainties are considered, the predictions show good agreement with the data.

\begin{table*}[!htbp]
  \caption{The number of observed dimuon events and the number of estimated events with statistical uncertainties in the full CDF \RNum{2} sample, corresponding to 9.4~\ifb of integrated luminosity, in five dimuon-mass ranges. The fractions for the individual SM processes are listed below the dashed line.}\label{table:background_plot_muon}
  \vspace{10pt}
  \centering
  \renewcommand{\arraystretch}{1.2}
  \begin{tabular}{c c c c c c c}
    \hline\hline
    \mmm ($\gevcc$)& $[40,60]$ & $[60,80]$ & $[80,100]$ & $[100,200]$ & $[200,350]$ \\
    \hline
    Data yield& $16754$& $18471$& $244729$& $13089$& $562$\\
    \hline
    Estimated yield& $15975\pm36$& $17786\pm30$& $244248\pm100$& $13184\pm24$& $580\pm5$\\ \myhdashline{6}
    $\zg\to\mm$& 91.45\%& 97.42\%& 99.86\%& 98.99\%& 97.78\%\\
    QCD multijet& 2.82\%& 0.54\%& 0.01\%& 0.09\%& 0.04\%\\
    $\zg\to\tau\tau$& 4.85\%& 1.40\%& 0.01\%& 0.06\%& 0.02\%\\
    Diboson& 0.42\%& 0.40\%& 0.12\%& 0.52\%& 1.32\%\\
    $W$+jets& 0.29\%& 0.07\%& 0.00\%& 0.04\%& 0.00\%\\
    $t\bar{t}$& 0.17\%& 0.16\%& 0.01\%& 0.31\%& 0.84\%\\
    \hline\hline
  \end{tabular}
\end{table*}

\begin{figure}
  \centering
  \includegraphics[width=0.5\textwidth]{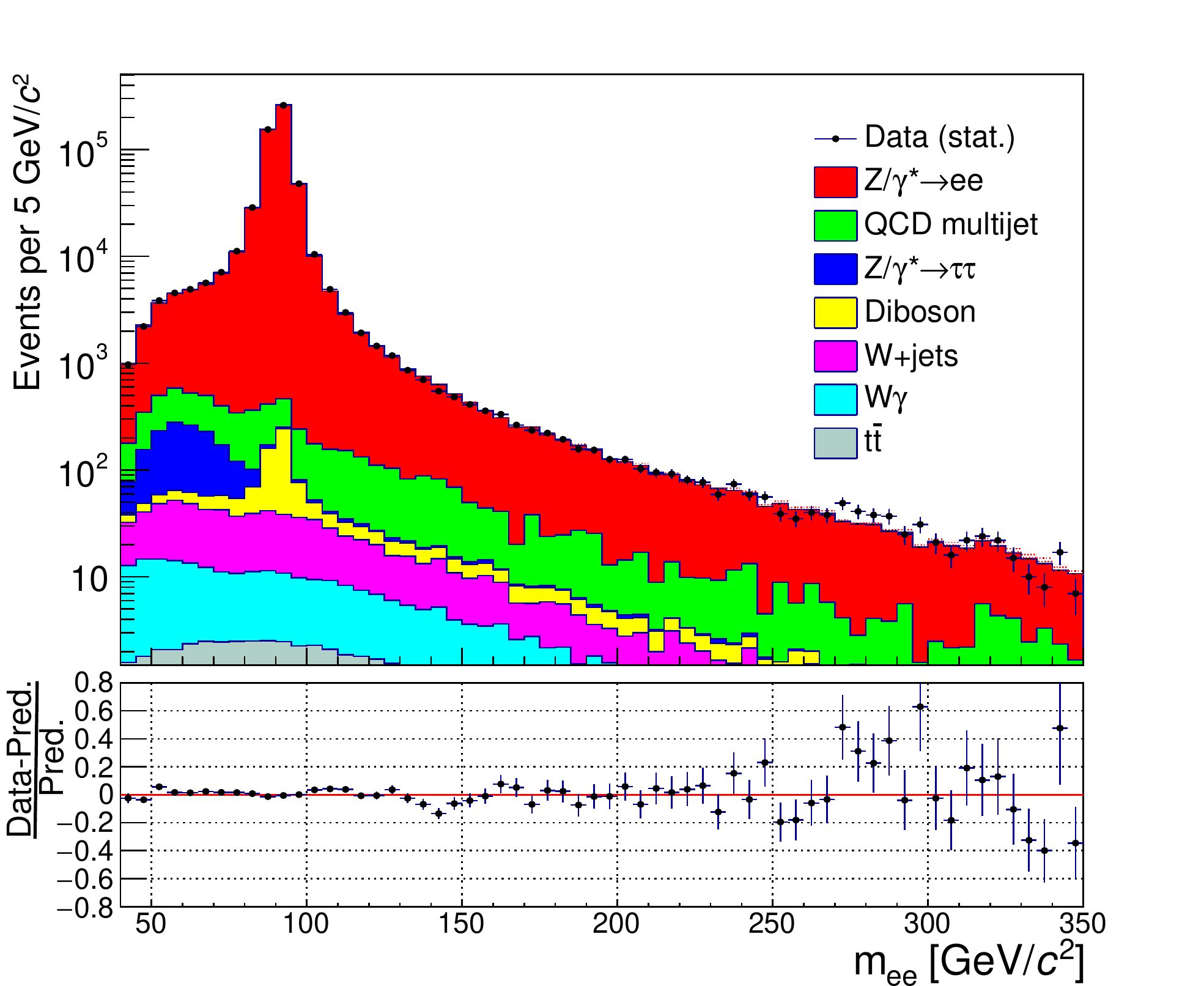}
  \caption{Observed dielectron mass distribution in data (circles) compared with the distributions from expected SM contributions (stacked). The normalized residuals of the data and the SM prediction are shown in the lower panel. Only statistical uncertainties are considered in both panels.}
  \label{fig:mee}
\end{figure}

For the electron channel, the $\zg \to ee$ signal process is the main contribution, and $\zg \to \tau\tau$, diboson ({\it WW}, {\it WZ}, {\it ZZ}), $t\bar{t}$, $W+$jets, $W\gamma$, and QCD multijet processes are considered as backgrounds. As in the case of the dimuon final state, all processes are modeled using simulation except for QCD multijet production, which is estimated from data.
The dielectron mass shape of QCD multijet events is obtained from data selected through inverse electron isolation requirements and normalized using a fit to the experimental dielectron mass distribution following Ref.~\cite{dyee}.
The sum of the expected SM processes agrees with the data as shown in Fig.~\ref{fig:mee} and Table~\ref{table:background_electron}.

\begin{table*}[!htbp]
  \caption{The number of observed dielectron events and the number of estimated events with statistical uncertainties in the full CDF \RNum{2} sample, corresponding to 9.4~\ifb of integrated luminosity, in five dielectron-mass ranges. The fractions for the individual SM processes are listed below the dashed line.}\label{table:background_electron}
  \vspace{10pt}
  \centering
  \renewcommand{\arraystretch}{1.2}
  \begin{tabular}{c c c c c c c}
    \hline\hline
    \mee ($\gevcc$) & $[40,60]$ & $[60,80]$ & $[80,100]$ & $[100,200]$ & $[200,350]$ \\
    \hline
    Data yield& $11590$& $28856$& $490211$& $27956$& $1357$\\
    \hline
    Estimated yield& $11416\pm53$& $28374\pm54$& $493391\pm158$& $27481\pm51$& $1307\pm15$\\ \myhdashline{6}
    $\zg\to ee$& 85.88\%& 93.78\%& 99.70\%& 94.67\%& 85.12\%\\
    QCD multijet& 7.58\%& 3.45\%& 0.18\%& 3.96\%& 10.98\%\\
    $\zg\to\tau\tau$& 4.70\%& 1.95\%& 0.01\%& 0.06\%& 0.12\%\\
    Diboson& 0.33\%& 0.22\%& 0.08\%& 0.35\%& 1.15\%\\
    $W$+jets& 1.03\%& 0.43\%& 0.02\%& 0.62\%& 1.48\%\\
    $W\gamma$& 0.43\%& 0.13\%& 0.01\%& 0.25\%& 0.87\%\\
    $t\bar{t}$& 0.07\%& 0.03\%& 0.00\%& 0.09\%& 0.28\%\\
    \hline\hline
  \end{tabular}
\end{table*}

\section{ISR Measurement in Drell-Yan Events}
\label{sec:isr}

The effect of the QCD ISR is probed by measuring the truncated mean of \zg \pt distribution in bins of the lepton-pair mass. Several corrections relevant for this measurement are described in this section.

First, the lepton energy and momentum are calibrated to correct for instrumental effects.
For the measurement using muon pairs, multiplicative and additive corrections are applied to the muon-track curvature as functions of the track pseudorapidity and azimuthal angle to correct inaccuracies of the magnetic field description and misalignment~\cite{dymm,rochester}.
The curvature resolution of the simulation is adjusted to match the data. For electron calibrations, the method used is similar, and multiplicative and additive corrections are applied to the electron energy as functions of data taking time, individual calorimeter tower, and the number of primary vertices~\cite{dyee}. The calorimeter resolution in simulation is also adjusted to agree with the data. 

The simulated $\ptll$ distribution is sensitive to the chosen model for the QCD ISR.
The parton-shower algorithm in {\sc Pythia6} uses a soft-collinear approximation, which is fast and efficient but is restricted in precision to the leading logarithmic order in the approximation used. This approach sums all emissions by evolving from large to small $Q^2$ by reformulating the DGLAP equations.  Uncertainties in this evolution and missing higher-order effects may introduce a mismodeling of the $\ptll$ distribution.
Mistuning of the intrinsic transverse momentum of the incoming partons may
also contribute to an inaccuracy.
Consequently, the simulated $\ptll$ distribution is adjusted to agree with the 
data by reweighting the \zg boson \pt at the generator level, where the correction is assumed to be independent of the lepton flavor.
This correction is iteratively extracted from the reconstructed $\ptll$ distributions
of both dielectron and dimuon events within the $Z$-peak mass region of 66--116 $\gevcc$.
The correction is a function of the \pt and the rapidity of the boson.
The \pt dependence of the weight is parametrized using a continuous piecewise polynomial function
of $\ln(\pt)$ and the rapidity dependence is parametrized using a linear function.

\begin{figure*}[!htbp]
  \includegraphics[width=0.49\textwidth]{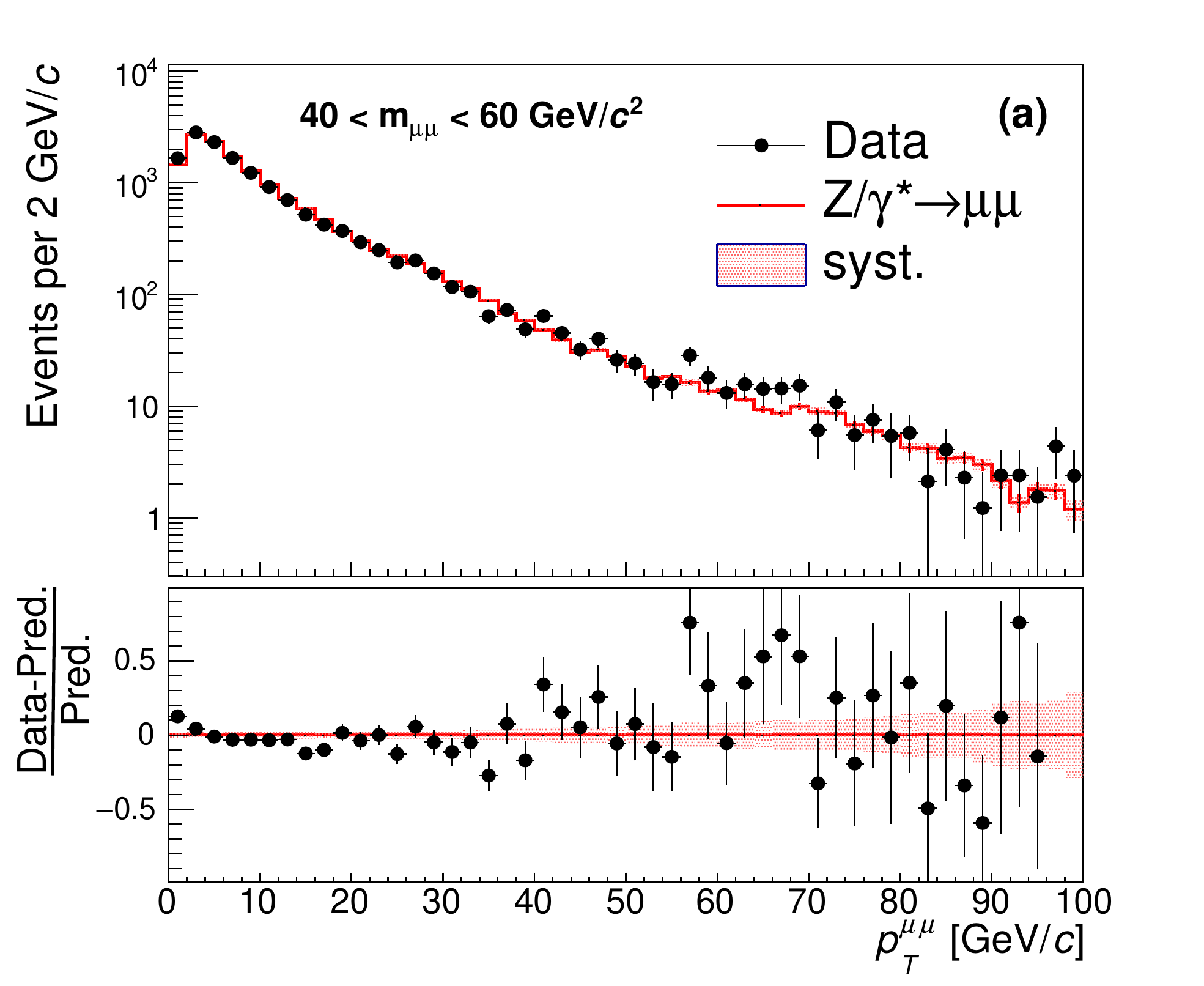}
  \includegraphics[width=0.49\textwidth]{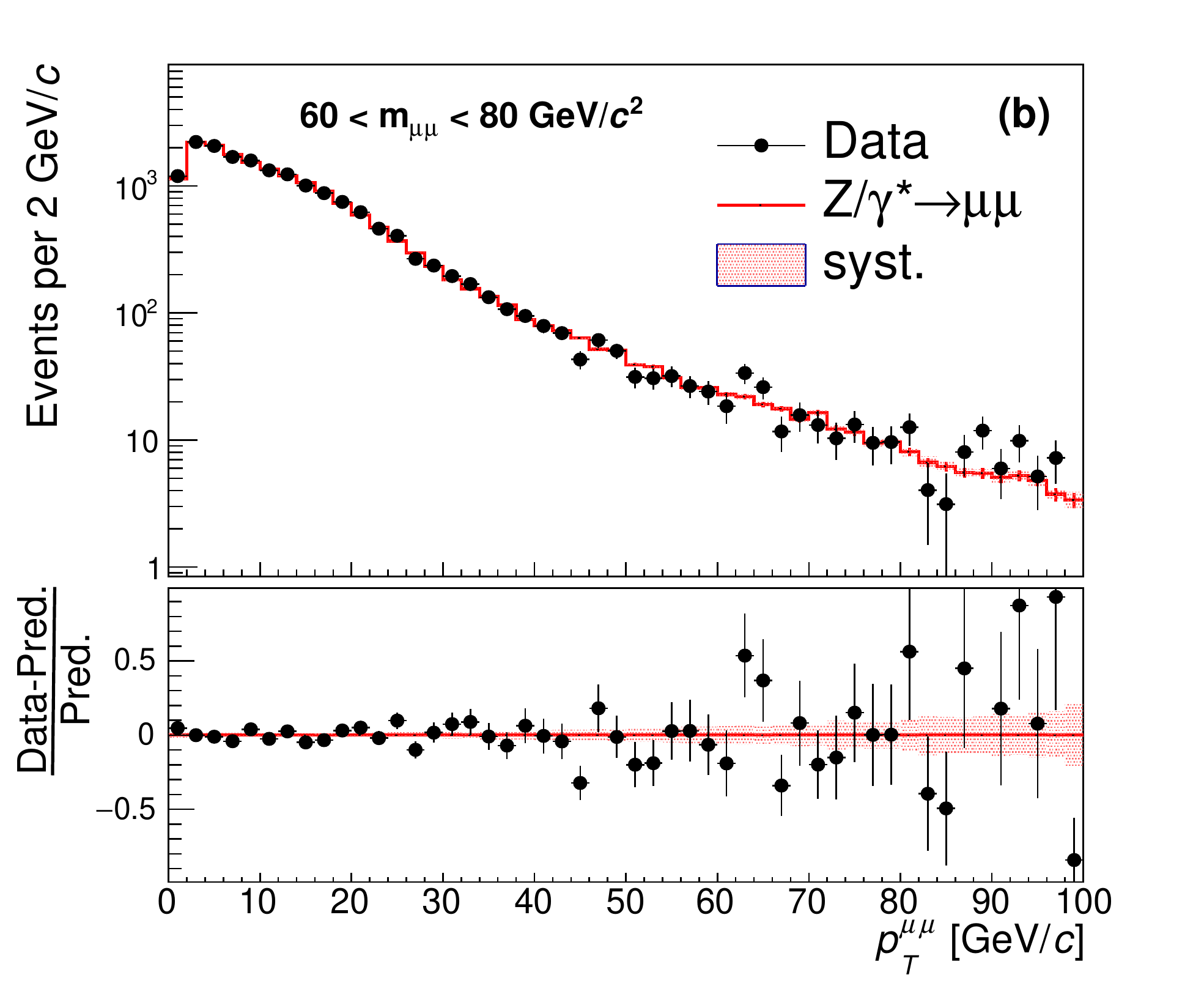}
  \includegraphics[width=0.49\textwidth]{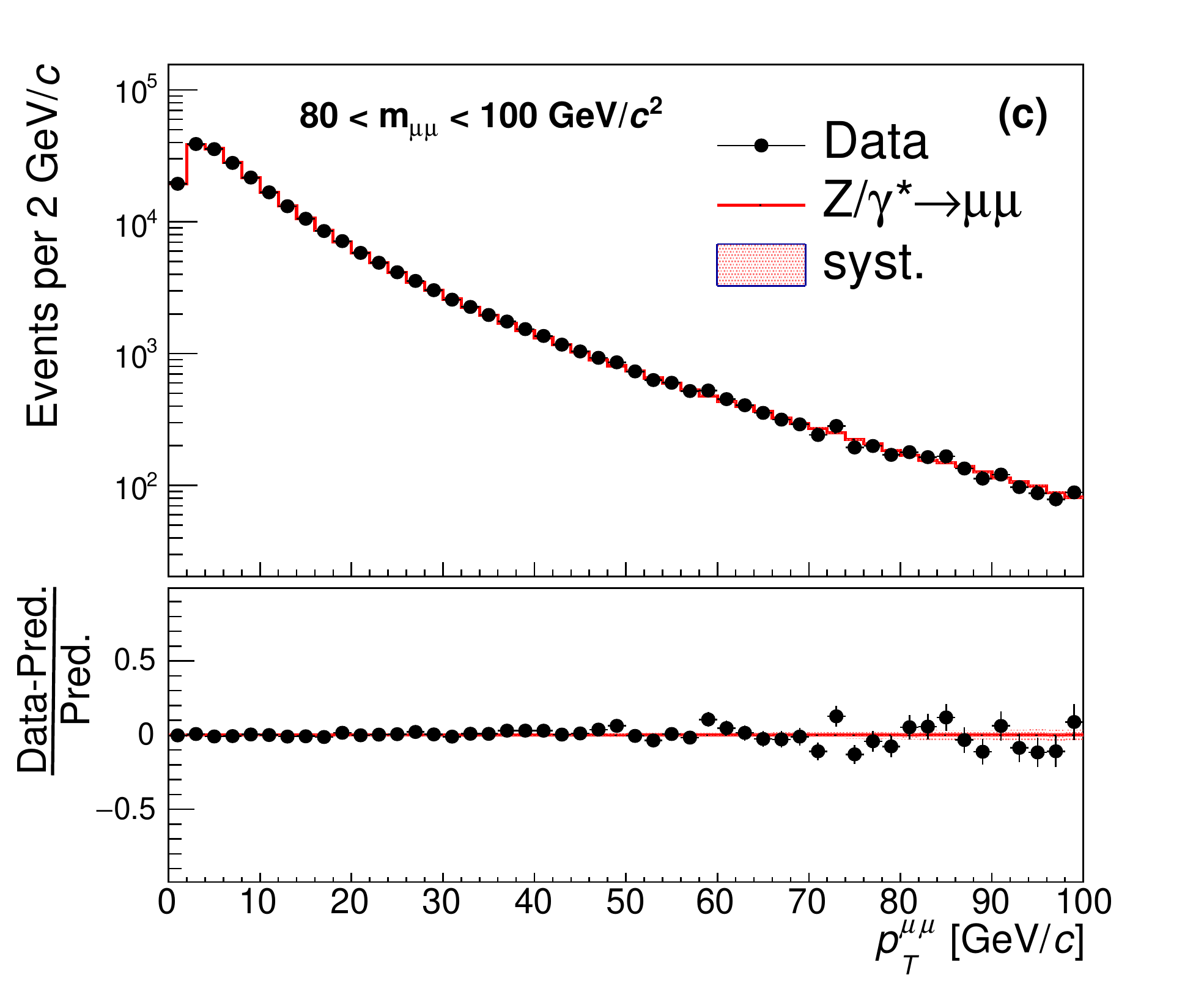}
  \includegraphics[width=0.49\textwidth]{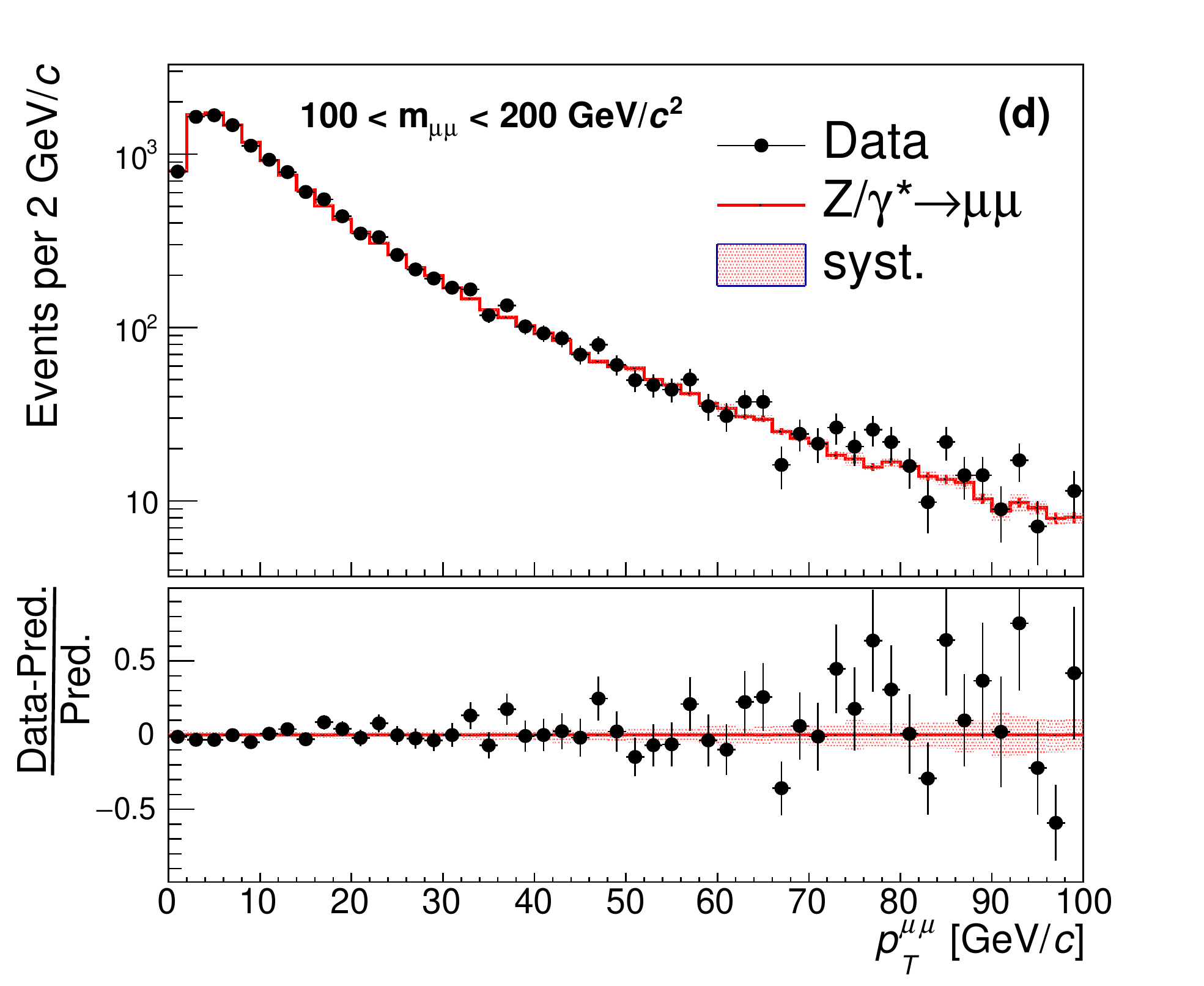}
  \includegraphics[width=0.49\textwidth]{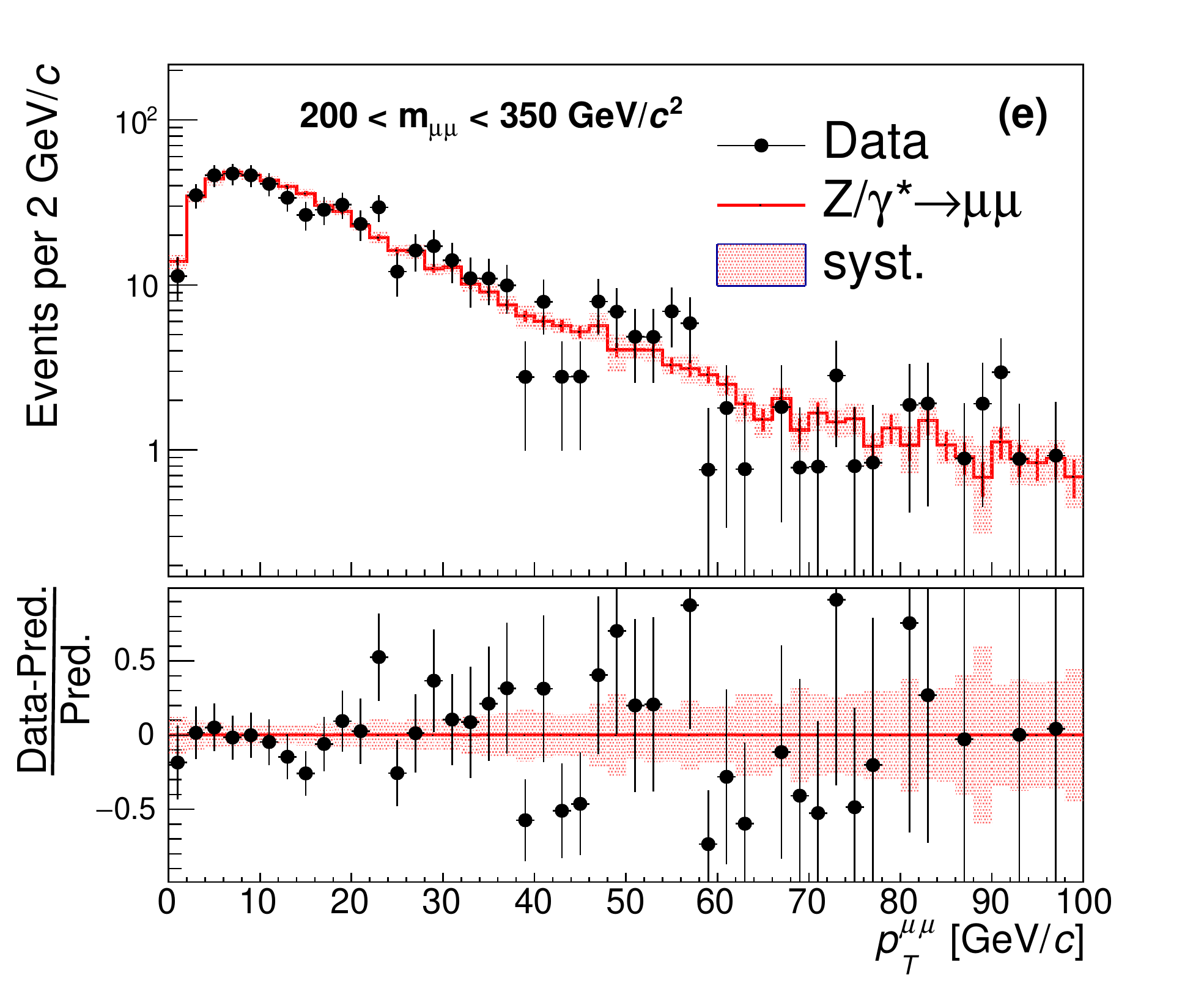}
  \caption{Distributions of the reconstructed dimuon $\pt$ in data (black circle) compared with distributions from the DY simulation after applying the $\zg$ boson $\pt$ corrections in each dimuon mass bin: (a) $[40~\gevcc,~60~\gevcc]$, (b) $[60~\gevcc,~80~\gevcc]$, (c) $[80~\gevcc,~100~\gevcc]$, (d) $[100~\gevcc,~200~\gevcc]$, (e) $[200~\gevcc,~350~\gevcc]$. Other backgrounds are subtracted from the data and the normalized residuals of the data and the DY simulation are shown in the lower panels. Statistical uncertainties (black bar) and systematic uncertainties (red shaded area) are shown.}
  \label{plot_ptmm}
\end{figure*}

\begin{figure*}[!htbp]
  \includegraphics[width=0.49\textwidth]{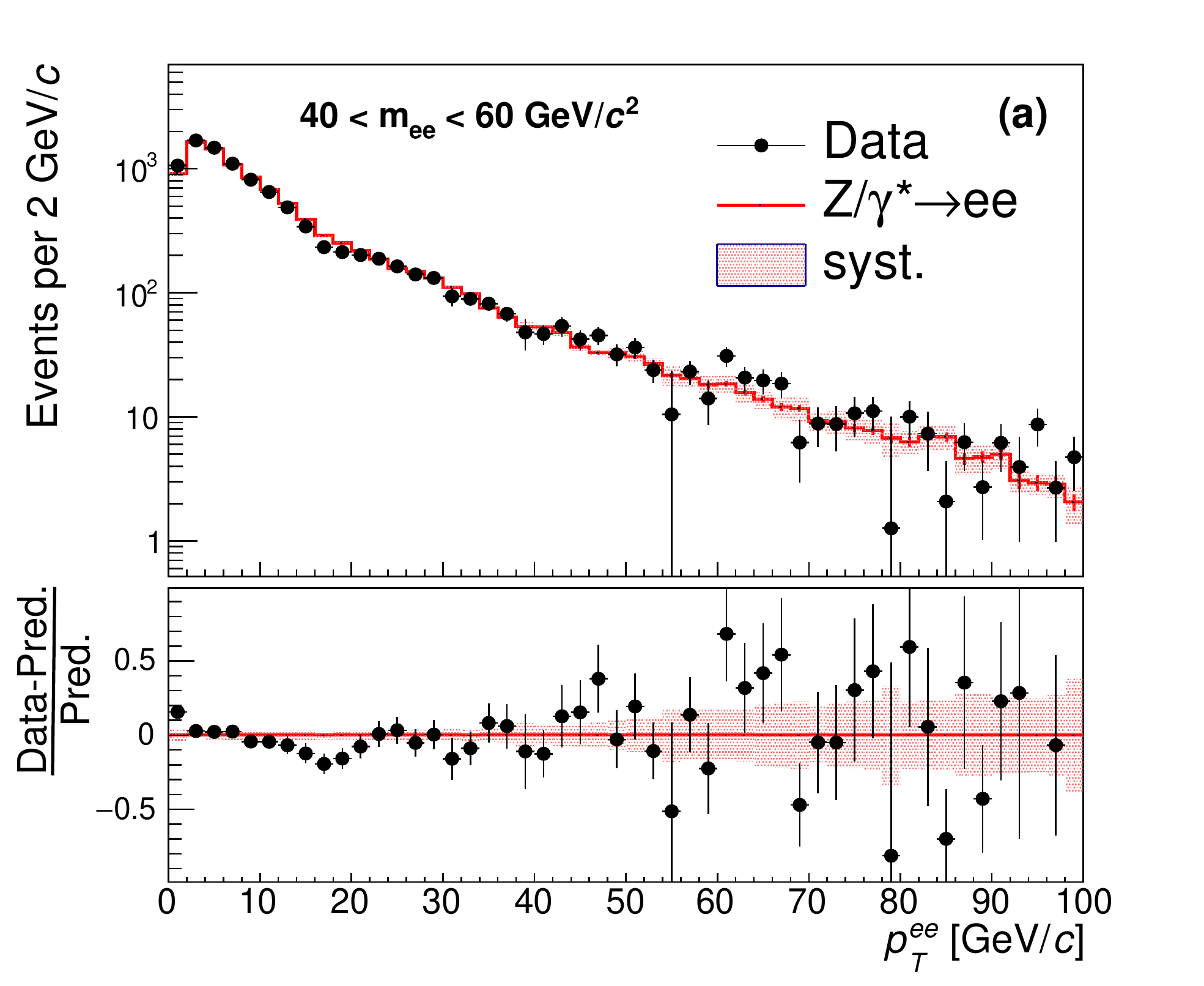}
  \includegraphics[width=0.49\textwidth]{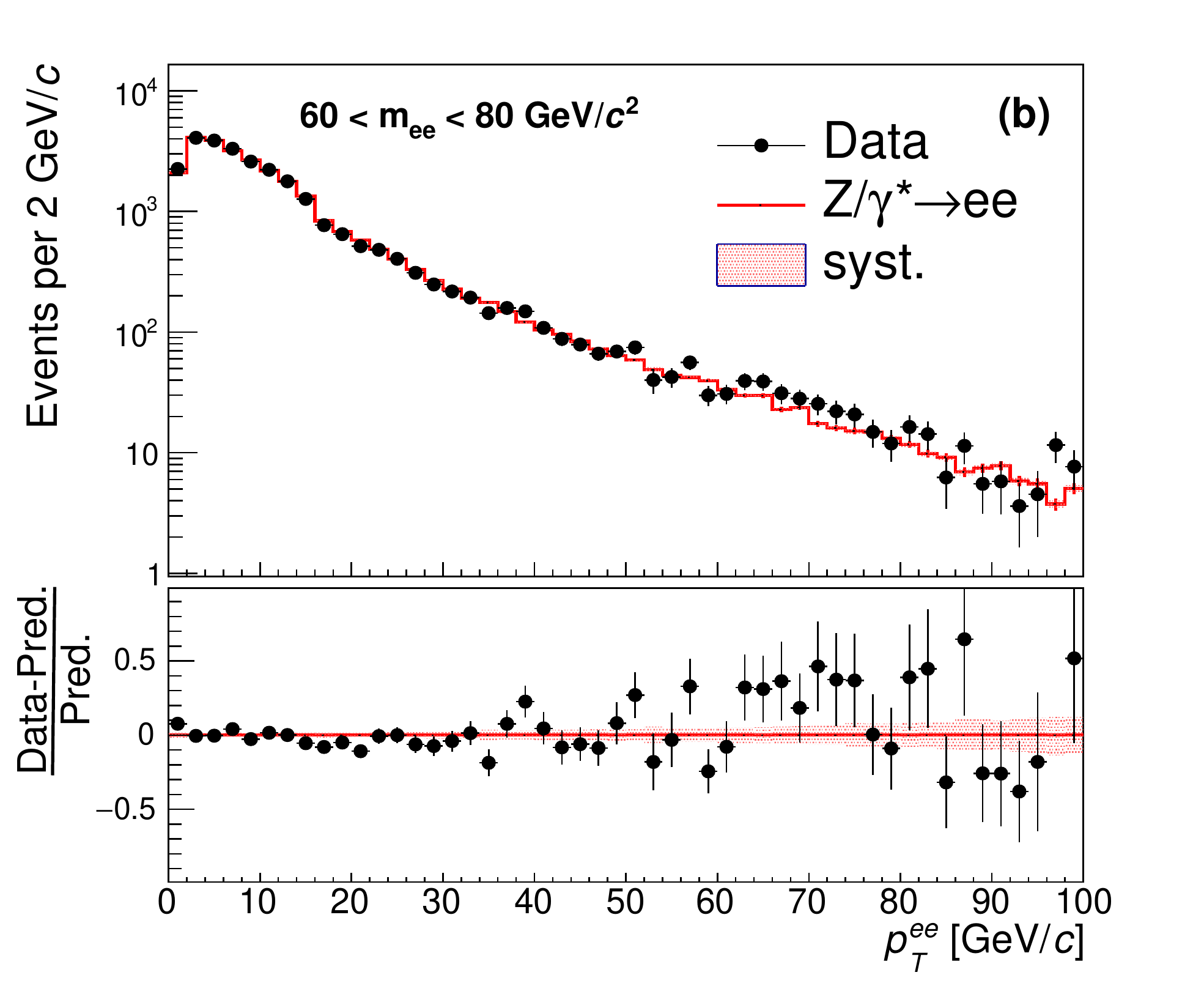}
  \includegraphics[width=0.49\textwidth]{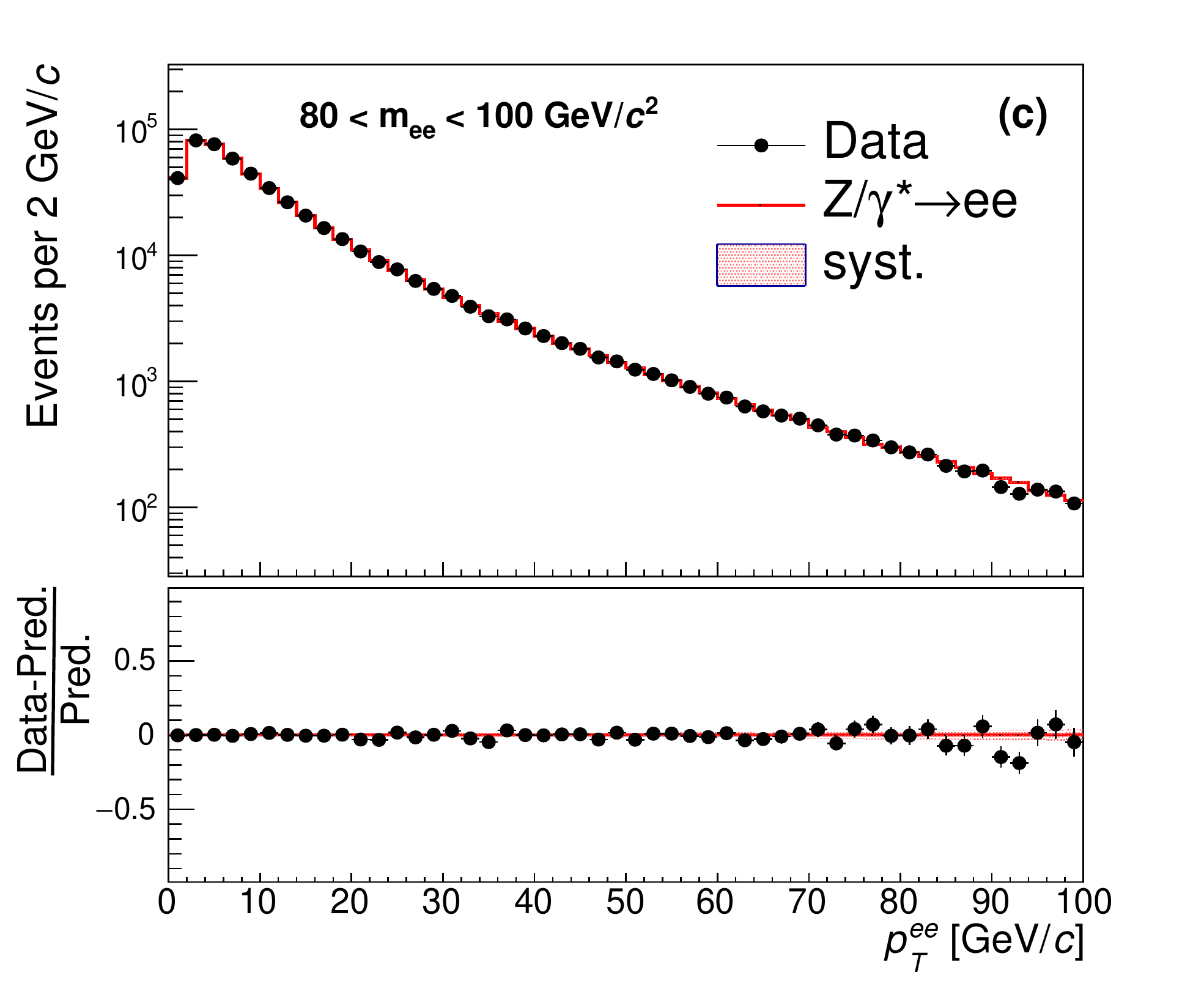}
  \includegraphics[width=0.49\textwidth]{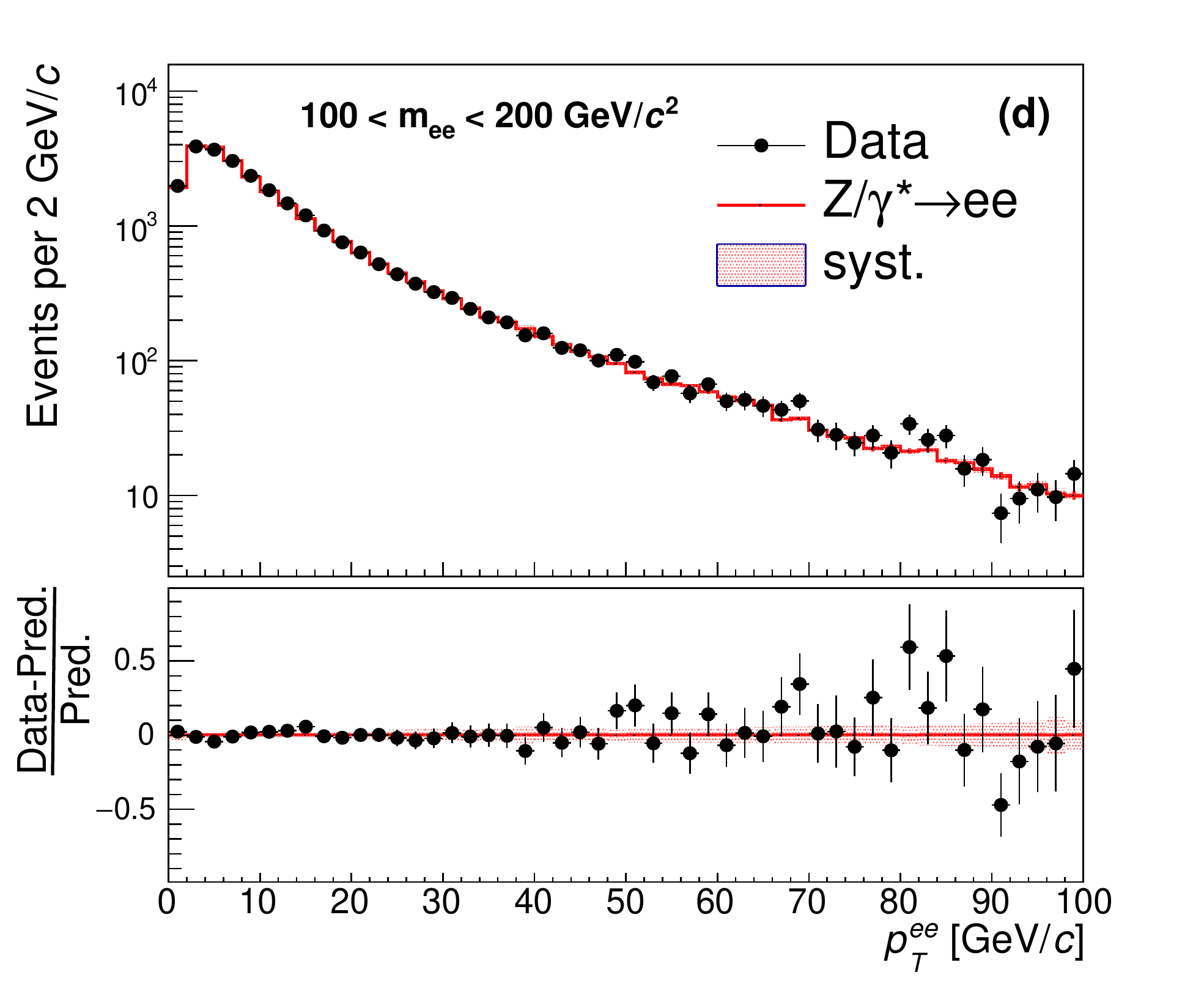}
  \includegraphics[width=0.49\textwidth]{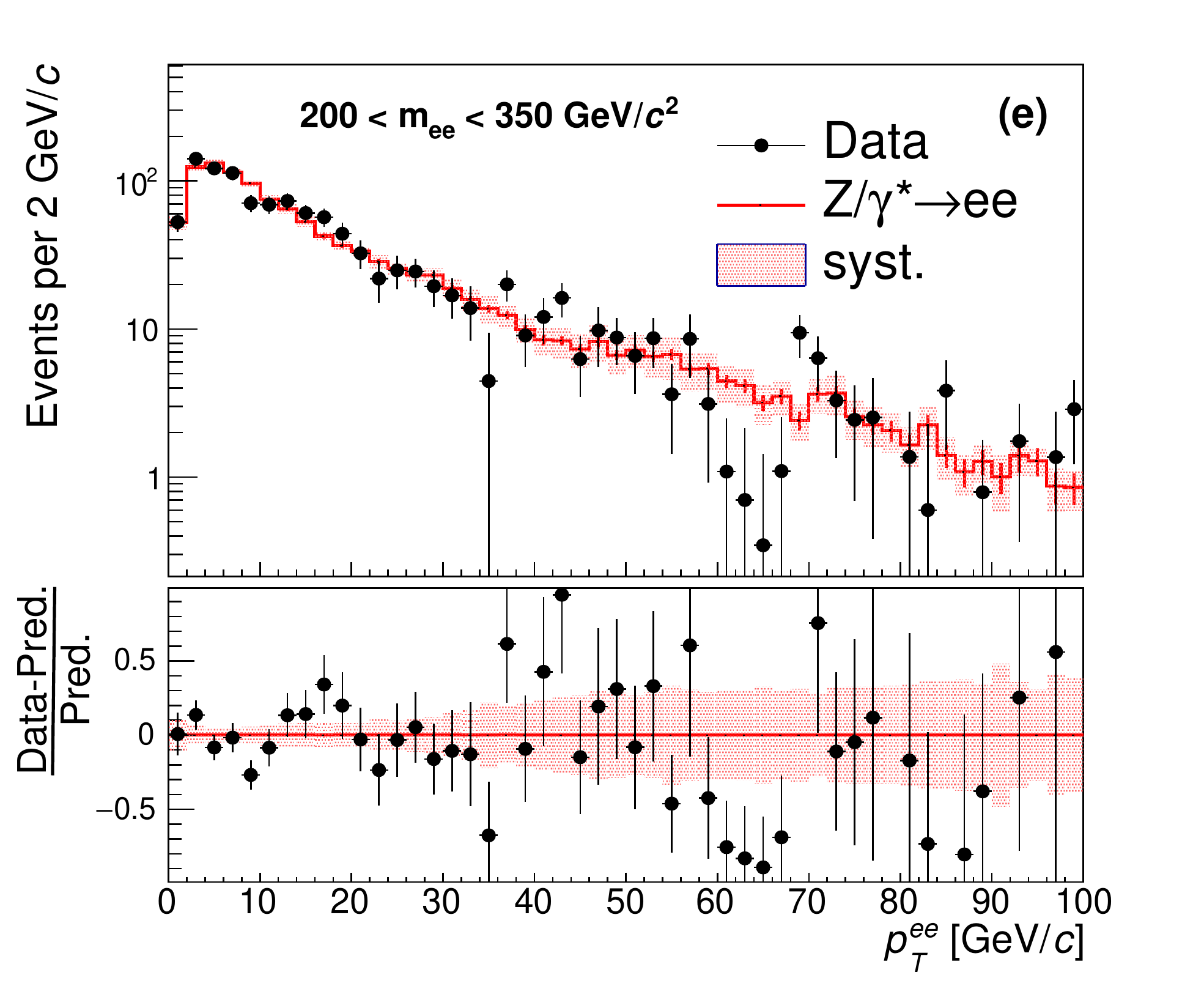}
  \caption{Distributions of the reconstructed dielectron $\pt$ in data (black circle) compared with distributions from the DY simulation after applying the $\zg$ boson $\pt$ corrections in each dielectron mass bin: (a) $[40~\gevcc,~60~\gevcc]$, (b) $[60~\gevcc,~80~\gevcc]$, (c) $[80~\gevcc,~100~\gevcc]$, (d) $[100~\gevcc,~200~\gevcc]$, (e) $[200~\gevcc,~350~\gevcc]$. Other backgrounds are subtracted from the data and the normalized residuals of the data and the DY simulation are shown in the lower panels. Statistical uncertainties (black bar) and systematic uncertainties (red shaded area) are shown.}
  \label{plot_ptee}
\end{figure*}

Fig.~\ref{plot_ptmm} and Fig.~\ref{plot_ptee} show the reconstructed $\ptll$ distribution of
the background-subtracted data and the simulated DY events after applying
the corrections described above. Good agreement is observed
between simulation and data for both $ee$ and $\mm$ final states.
 
The mean values of the \zg transverse momentum \ptzge in DY events are obtained from the average values of the reconstructed \ptll distribution \ptlle by multiplying corrections for acceptance, detector and QED FSR effects.
The mean values of the \zg mass $\mzge$ are also obtained from the average values of the reconstructed lepton-pair mass $\mlle$ by the same procedure.
Combined corrections for the average \pt and mass of the lepton pairs within a
mass bin, denoted by $\mathcal{R}_{\pt}$ and $\mathcal{R}_{\rm m}$ respectively,
are derived from the simulated DY sample and applied as shown below,

\begin{linenomath}
\begin{equation} \label{eq:pt}
  \begin{aligned}[t]
    \ptzge_{\mathrm {data}} &= \mathcal{R}_{\pt} \times \ptlle_{\mathrm{data}} \\
    \text{where } \mathcal{R}_{\pt} & \equiv \frac{\ptzge^{\mathrm{gen}}_{\mathrm{MC}}} {\ptlle^{\mathrm{det}}_{\mathrm{MC}}},
  \end{aligned}
\end{equation}
\begin{equation} \label{eq:m}
  \begin{aligned}[t]
    {\langle \mzg \rangle}_{\mathrm{data}} &= \mathcal{R}_{\rm m}\times {\langle \mll \rangle}_{\mathrm{data}} \\
    \text{where } \mathcal{R}_{\rm m} & \equiv\frac{{\langle \mzg \rangle}^{\mathrm{gen}}_{\mathrm{MC}}}{{\langle \mll \rangle}^{\mathrm{det}}_{\mathrm{MC}}}.
  \end{aligned}
\end{equation}
\end{linenomath}

Here, $\ptzge^{\mathrm{gen}}_{\mathrm{MC}}$ and
$\langle \mzg \rangle^{\mathrm{gen}}_{\mathrm{MC}}$ are generator-level averages over the full phase space within a mass bin,
and $\ptlle^{\mathrm{det}}_{\mathrm{MC}}$ and
$\langle \mll \rangle^{\mathrm{det}}_{\mathrm{MC}}$ are detector-level
averages corresponding to the data.
The corrections for \pt ($\mathcal{R}_{\pt}$) range from 0.75 to 1.00 and the corrections for $m$ ($\mathcal{R}_{\rm m}$) range from 0.82 to 1.01.

\section{Systematic Uncertainties}
\label{sec:uncertainty}

The QCD ISR model, QED FSR model, energy and momentum corrections on leptons, and background normalizations are sources of systematic uncertainty of the measurements of $\ptzge$ and $\mzge$. The uncertainties of $\ptzge$ for the dimuon and dielectron final states are listed in Tables~\ref{tab:mu_sys} and \ref{tab:el_sys}, respectively. The QCD ISR model and QED FSR model uncertainties dominate.

The QCD ISR uncertainty is defined to be the uncertainty of the underlying $\zg$-boson \pt distribution used in the DY simulation.
The simulated \zg boson \ptll distribution is improved by
reweighting the \pt distribution so as to agree with the data.
Two sources of uncertainty of this correction are considered: 
the correction derived from events in the $Z$-mass region
to mass values away from the $Z$ peak, and the statistical uncertainty to
the correction. The QCD ISR uncertainty is estimated using a set of
simulated experiments in which the input values are varied within uncertainties from both sources.

The reconstructed $\zg$ mass and \pt can be affected by the presence of QED FSR which is not fully contained in the reconstructed event products.
The simulation accounts for this in the correction factors $\mathcal{R}_{\pt}$ and $\mathcal{R}_{\rm m}$.
The default model of QED FSR in {\sc Pythia6} is augmented with a QED shower algorithm which could be inaccurate for hard QED emissions.
The {\sc Photos} QED generator~\cite{photos} calculates single-photon radiation in the leading order of the electromagnetic interaction and simulates multiphoton
radiation by iterating the single-photon calculation.
To estimate the QED FSR uncertainty to the correction factors,
the generator-level results for $\ptzge$ and $\mzge$ before and after
QED FSR are evaluated for {\sc Pythia6} and {\sc Photos}.
The difference is taken as the QED FSR uncertainty.

The systematic uncertainty due to the electron energy and muon momentum scales is
estimated using simulated samples by varying the scales within their uncertainties.
The systematic uncertainty due to simulation inaccuracies in the energy and momentum resolution is evaluated by varying the simulation parameters that control the resolution smearing.

The uncertainties from the simulation-based description of the background processes are estimated by changing each cross section by its uncertainty, where the factorization, renormalization, and PDF uncertainties are considered. The 6\% uncertainty on the determination of the integrated luminosity~\cite{dyee} is included. The background normalizations are treated as 100\% correlated.

Using averages of $\mzg$ over large bins may lead to biased results when representing $\ptzge$ as a function of $\mzg$~\cite{binning}. The additional systematic uncertainties accounting for this binning effect are considered for \mzge measurement by taking the difference between $\ln \mzgsqe$ and $\langle \ln \mzgsq \rangle$. This is the dominant systematic uncertainty for the $\mzge$ measurement, and ranges from 0.04\% to 1.13\%.

\begin{table*}[htbp]
  \caption{Fractional statistical and systematic uncertainties of \ptzge (in \%) for each dimuon mass bin. Individual sources of systematic uncertainties are listed below the dashed line.} 
  \label{tab:mu_sys}
  \vspace{10pt}
  \centering
  \renewcommand{\arraystretch}{1.2}
  \begin{tabular}{l c c c c c}
    \hline\hline                      
    Mass bin ($\gevcc$) & $[40,60]$ & $[60,80]$ & $[80,100]$ & $[100,200]$ & $[200,350]$ \\
    \hline
    Statistical uncertainty (\%)&      0.96&      0.73&      0.22&      0.95&      3.79\\
    \hline
    Systematic uncertainty (\%)&      1.29&      1.33&      0.26&      0.90&      2.76\\ \myhdashline{6}
    \,   ISR model (\%)&      0.90&      0.93&      0.24&      0.50&      2.28\\
    \,   QED FSR model (\%)&      0.87&      0.93&      0.03&      0.18&      0.41\\
    \,   Momentum scale (\%)&      0.12&      0.13&      0.04&      0.20&      0.86\\
    \,   Momentum resolution (\%)&      0.07&      0.06&      0.08&      0.68&      1.20\\
    \,   Background normalization (\%)&      0.28&      0.10&      0.03&      0.16&      0.25\\
\hline\hline
  \end{tabular}
\end{table*}

\begin{table*}[htbp]
  \caption{Fractional statistical and systematic uncertainties of \ptzge (in \%) for each dielectron mass bin. Individual sources of systematic uncertainties are listed below the dashed line.} 
  \label{tab:el_sys}
  \vspace{10pt}
  \centering
  \renewcommand{\arraystretch}{1.2}
  \begin{tabular}{l c c c c c}
    \hline\hline    
    Mass bin ($\gevcc$) & $[40,60]$ & $[60,80]$ & $[80,100]$ & $[100,200]$ & $[200,350]$ \\
    \hline
    Statistical uncertainty (\%)&      1.38&      0.70&      0.16&      0.72&      3.44\\
    \hline
    Systematic uncertainty (\%)&      2.03&      0.89&      0.12&      0.69&      2.11\\
    \myhdashline{6}
    \,   ISR model (\%)&      1.37&      0.47&      0.06&      0.33&      1.68\\
    \,   QED FSR model (\%)&      1.39&      0.72&      0.05&      0.22&      0.67\\
    \,   Energy scale (\%)&      0.13&      0.08&      0.02&      0.07&      0.19\\
    \,   Energy resolution (\%)&      0.02&      0.02&      0.08&      0.10&      0.15\\
    \,   Background normalization (\%)&      0.57&      0.21&      0.03&      0.56&      1.06\\
    \hline\hline
  \end{tabular}
\end{table*}

\section{Results}
\label{sec:result}

\begin{figure}
  \centering
  \includegraphics[width=0.5\textwidth]{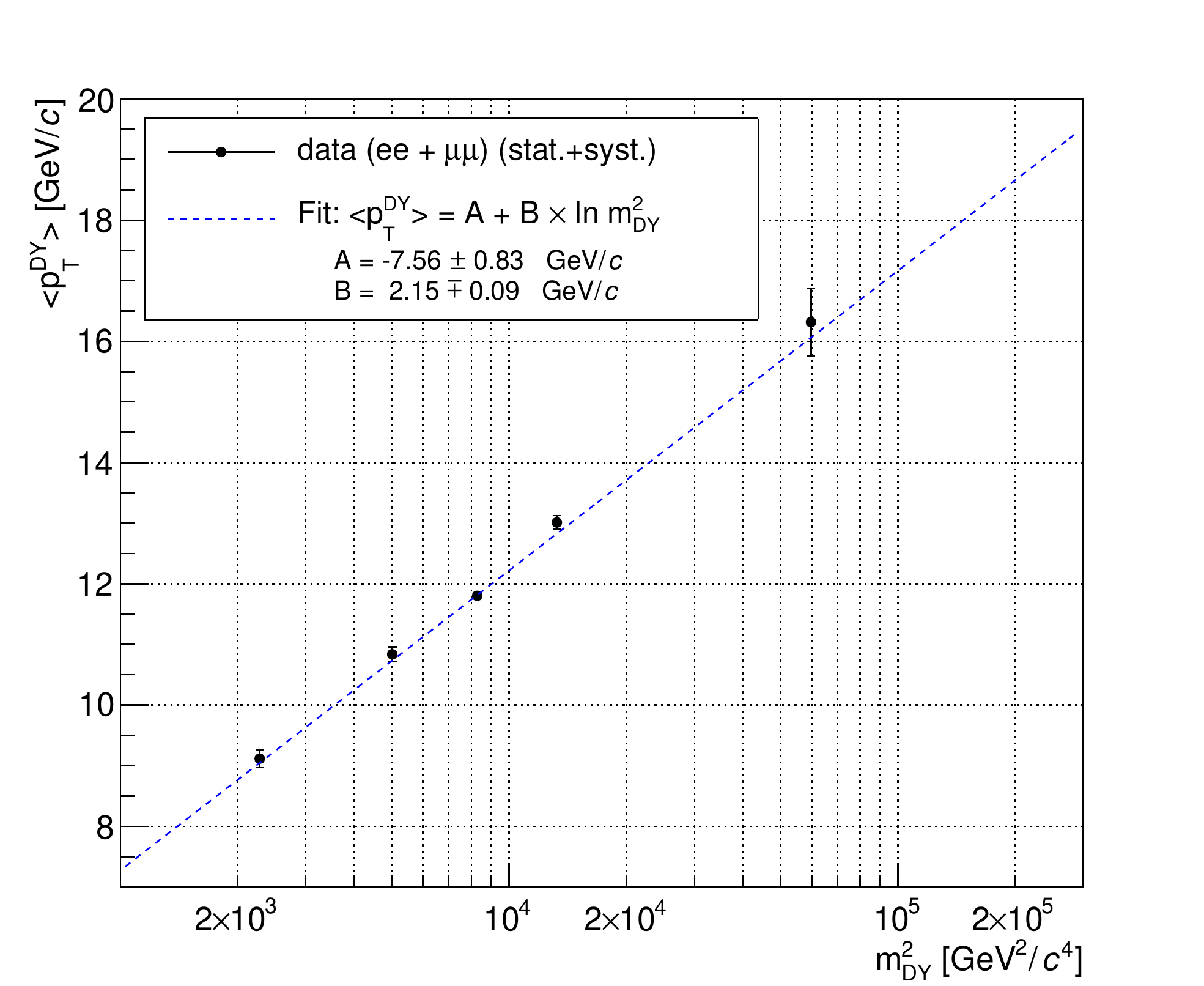}
  \caption{Average transverse momentum of \zg ($\ptzg$) as a function of $\ln\mzgsq$ fitted with a linear function. Statistical and systematic uncertainties are included in the results of the fit.}
  \label{plot_result_graph_comb}
\end{figure}

The measurements of $\ptzge$ and ${\langle \mzg \rangle}$ in DY events are obtained for dimuon and dielectron final states. The dimuon and dielectron results are in good agreement as shown in Table~\ref{table:result_mm} and Table~\ref{table:result_ee}. The results are combined using the best linear unbiased estimation~\cite{combine} and are given in Table~\ref{table:result_comb}.
Systematic uncertainties are combined accounting for all the correlations
between final states and are listed in Table~\ref{table:sys_comb}.
Fig.~\ref{plot_result_graph_comb} shows the dependence of $\ptzge$ as a function of $\ln \mzgsq$, which is modeled by the linear function 
\begin{linenomath}
\begin{equation}
\ptzge =
   (-7.56\pm0.83) + (2.15\mp0.09) \ln \mzgsq~[\gevc],
\end{equation}
\end{linenomath}
where \mzg is the value of the \zg mass measured in units of $\gevcc$. The average symbol for \mzg is omitted as the additional systematic uncertainty for the binning effect is included. The two fitted parameters are found to be strongly anti-correlated (-99.97\%)\footnote{One can reduce the correlation between the two parameters for easier interpretation by using a modified fitting function $\ptzge = (11.82\pm0.02) + (2.15\pm0.09) \ln {\mzgsq \over \mzpolesq}~[\gevc]$, where \mzpole is 91.19~$\gevcc$. In this case, the two parameters are positively correlated by approximately 12\%.}.

\begin{table*}[hp]
  \caption{Average values and their uncertainties of the dimuon mass and $\pt$ for $\mm$ events.}
  \label{table:result_mm}
  \vspace{10pt}
  \centering
  \renewcommand{\arraystretch}{1.2}
  \begin{tabular}{c p{0.5cm} r p{0.5cm} r}
    \hline\hline
    \parbox{2cm}{\centering Mass bin\\($\gevcc$)} & & \parbox{4cm}{\centering $\mzge{\pm}{\rm stat}{\pm}{\rm syst}$\\($\gevcc$)} & & \parbox{4cm}{\centering $\ptzge{\pm}{\rm stat}{\pm}{\rm syst}$\\($\gevcc$)} \rule{0pt}{4ex}\\\hline
           $[40,60]$&&       $47.72\pm0.05\pm0.31$&&       $9.12\pm0.09\pm0.12$\\
           $[60,80]$&&       $70.66\pm0.04\pm0.27$&&       $10.81\pm0.08\pm0.14$\\
          $[80,100]$&&       $90.99\pm0.01\pm0.08$&&       $11.84\pm0.03\pm0.03$\\
         $[100,200]$&&       $115.29\pm0.18\pm1.30$&&      $13.17\pm0.12\pm0.12$\\
         $[200,350]$&&       $243.33\pm1.63\pm2.52$&&      $16.18\pm0.61\pm0.45$\\
\hline\hline
\end{tabular}
\end{table*}

\begin{table*}[hp]
  \caption{Average values and their uncertainties of the dielectron mass and $\pt$ for $ee$ events.}
  \label{table:result_ee}
  \vspace{10pt}
  \centering
  \renewcommand{\arraystretch}{1.2}
  \begin{tabular}{c p{0.5cm} r p{0.5cm} r}
    \hline\hline
    \parbox{2cm}{\centering Mass bin\\($\gevcc$)} & & \parbox{4cm}{\centering $\mzge{\pm}{\rm stat}{\pm}{\rm syst}$\\($\gevcc$)} & & \parbox{4cm}{\centering $\ptzge{\pm}{\rm stat}{\pm}{\rm syst}$\\($\gevcc$)} \rule{0pt}{4ex}\\\hline
           $[40,60]$&&         $47.83\pm0.05\pm0.31$&&          $9.10\pm0.13\pm0.18$\\
           $[60,80]$&&         $70.76\pm0.04\pm0.26$&&         $10.84\pm0.08\pm0.10$\\
          $[80,100]$&&         $90.98\pm0.01\pm0.08$&&         $11.79\pm0.02\pm0.01$\\
         $[100,200]$&&        $115.11\pm0.13\pm1.31$&&         $12.93\pm0.09\pm0.09$\\
         $[200,350]$&&        $245.46\pm1.29\pm2.61$&&         $16.41\pm0.56\pm0.35$\\
\hline\hline
\end{tabular}
\end{table*}

\begin{table*}[hp]
  \caption{Average values and their uncertainties of the lepton-pair mass and $\pt$, combining $ee$ and $\mm$ events.}
  \label{table:result_comb}
  \vspace{10pt}
  \centering
  \renewcommand{\arraystretch}{1.2}
  \begin{tabular}{c p{0.5cm} r p{0.5cm} r}
    \hline\hline
    \parbox{2cm}{\centering Mass bin\\($\gevcc$)} & & \parbox{4cm}{\centering $\mzge{\pm}{\rm stat}{\pm}{\rm syst}$\\($\gevcc$)} & & \parbox{4cm}{\centering $\ptzge{\pm}{\rm stat}{\pm}{\rm syst}$\\($\gevcc$)} \rule{0pt}{4ex}\\\hline
           $[40,60]$&&         $47.77\pm0.04\pm0.31$&&          $9.12\pm0.09\pm0.12$\\
           $[60,80]$&&         $70.75\pm0.03\pm0.26$&&         $10.84\pm0.07\pm0.10$\\
          $[80,100]$&&         $90.98\pm0.00\pm0.08$&&         $11.80\pm0.02\pm0.02$\\
         $[100,200]$&&        $115.19\pm0.11\pm1.30$&&         $13.01\pm0.07\pm0.09$\\
         $[200,350]$&&        $244.49\pm1.02\pm2.56$&&         $16.32\pm0.42\pm0.36$\\
\hline\hline
\end{tabular}
\end{table*}

\begin{table*}[hp]
  \caption{Statistical and systematic uncertainties of \ptzge in the combined dielectron and dimuon final states. Fractional uncertainties are shown for the various dilepton mass bins and each systematic source.}
  \label{table:sys_comb}
  \vspace{10pt}
  \centering
  \renewcommand{\arraystretch}{1.2}
  \begin{tabular}{l r r r r r}
    \hline\hline
    Mass bin ($\gevcc$) &   [40,60]&   [60,80]&  [80,100]& [100,200]& [200,350]\\\hline
    Statistical uncertainty (\%)&      0.96&      0.62&      0.14&      0.57&      2.56\\
    [0.5ex]\hline
    Systematic uncertainty (\%)&      1.29&      0.93&      0.13&      0.66&      2.23\\
    \myhdashline{6}
    \,   ISR model (\%)&      0.90&      0.52&      0.09&      0.39&      1.93\\
    \,   QED FSR model (\%)&      0.87&      0.74&      0.05&      0.20&      0.56\\
    \,  Muon momentum scale (\%)&      0.12&      0.01&      0.01&      0.07&      0.35\\
    \,  Electron energy scale (\%)&      0.00&      0.07&      0.01&      0.05&      0.11\\
    \,  Muon momentum resolution (\%)&      0.07&      0.01&      0.02&      0.22&      0.49\\
    \,   Electron energy resolution (\%)&      0.00&      0.02&      0.07&      0.07&      0.09\\
    \,   Background normalization (\%)&      0.28&      0.20&      0.03&      0.43&      0.73\\
    \hline\hline
  \end{tabular}
\end{table*}

\section{Conclusions}
\label{sec:conclusion}

This analysis presents a novel approach to characterize the effect of QCD ISR and a measurement of the ISR activity in DY events produced in proton-antiproton
collisions at $\sqrt{s}~=~1.96$ TeV, corresponding to an integrated luminosity of 9.4~fb$^{-1}$.
The measurement of the average lepton-pair \pt distribution is performed as a function of the dilepton invariant mass to quantify the effect of ISR activity.
The average value of the lepton-pair $\pt$ is found to have a linear dependence on $\ln \mzgsq$. The dependence is described by the function,
$\ptzge = A + B \ln \mzgsq $, where the parameters A and B are obtained by a linear fit to be $-7.56 \pm 0.83~\gevc$ and $2.15 \mp 0.09~\gevc$, respectievly.
The linear dependence is an original finding of this work and may be exploited to improve the modeling of ISR. Since there is no theoretical prediction of this measurement, theorists may usefully make calculations of this measurement under various models of ISR. In addition, by means of a linear extrapolation this finding can be used to estimate the expected ISR activity at higher masses and is relevant to searches for BSM particles.

%
%

\begin{acknowledgements}
  This document was prepared by the CDF collaboration using the resources of the
  Fermi National Accelerator Laboratory (Fermilab), a U.S.  Department of Energy,
  Office of Science, HEP User Facility. Fermilab is managed by Fermi Research
  Alliance, LLC (FRA), acting under Contract No.  DE-AC02-07CH11359.  We thank
  the Fermilab staff and the technical staffs of the participating institutions
  for their vital contributions. This work was supported by the U.S. Department
  of Energy and National Science Foundation; the Italian Istituto Nazionale di
  Fisica Nucleare; the Ministry of Education, Culture, Sports, Science and
  Technology of Japan; the Natural Sciences and Engineering Research Council of
  Canada; the National Science Council of the Republic of China; the Swiss
  National Science Foundation; the A.P. Sloan Foundation; the Bundesministerium
  f\"ur Bildung und Forschung, Germany; the National Research Foundation of Korea
  and the Promising-Pioneering Researcher Program through Seoul National University;
  the Science and Technology Facilities Council
  and the Royal Society, United Kingdom; the Russian Foundation for Basic
  Research; the Ministerio de Ciencia e Innovaci\'{o}n, and Programa
  Consolider-Ingenio 2010, Spain; the Slovak R\&D Agency; the Academy of Finland;
  and the Australian Research Council (ARC).
\end{acknowledgements}



\end{document}

%% file: author_prd.tex
\affiliation{Institute of Physics, Academia Sinica, Taipei, Taiwan 11529, Republic of China}
\affiliation{Argonne National Laboratory, Argonne, Illinois 60439, USA}
\affiliation{University of Athens, 157 71 Athens, Greece}
\affiliation{Institut de Fisica d'Altes Energies, ICREA, Universitat Autonoma de Barcelona, E-08193, Bellaterra (Barcelona), Spain}
\affiliation{Baylor University, Waco, Texas 76798, USA}
\affiliation{Istituto Nazionale di Fisica Nucleare Bologna, \ensuremath{^{kk}}University of Bologna, I-40127 Bologna, Italy}
\affiliation{University of California, Davis, Davis, California 95616, USA}
\affiliation{University of California, Los Angeles, Los Angeles, California 90024, USA}
\affiliation{Instituto de Fisica de Cantabria, CSIC-University of Cantabria, 39005 Santander, Spain}
\affiliation{Carnegie Mellon University, Pittsburgh, Pennsylvania 15213, USA}
\affiliation{Enrico Fermi Institute, University of Chicago, Chicago, Illinois 60637, USA}
\affiliation{Comenius University, 842 48 Bratislava, Slovakia; Institute of Experimental Physics, 040 01 Kosice, Slovakia}
\affiliation{Joint Institute for Nuclear Research, RU-141980 Dubna, Russia}
\affiliation{Duke University, Durham, North Carolina 27708, USA}
\affiliation{Fermi National Accelerator Laboratory, Batavia, Illinois 60510, USA}
\affiliation{University of Florida, Gainesville, Florida 32611, USA}
\affiliation{Laboratori Nazionali di Frascati, Istituto Nazionale di Fisica Nucleare, I-00044 Frascati, Italy}
\affiliation{University of Geneva, CH-1211 Geneva 4, Switzerland}
\affiliation{Glasgow University, Glasgow G12 8QQ, United Kingdom}
\affiliation{Harvard University, Cambridge, Massachusetts 02138, USA}
\affiliation{Division of High Energy Physics, Department of Physics, University of Helsinki, FIN-00014, Helsinki, Finland; Helsinki Institute of Physics, FIN-00014, Helsinki, Finland}
\affiliation{University of Illinois, Urbana, Illinois 61801, USA}
\affiliation{The Johns Hopkins University, Baltimore, Maryland 21218, USA}
\affiliation{Institut f\"{u}r Experimentelle Kernphysik, Karlsruhe Institute of Technology, D-76131 Karlsruhe, Germany}
\affiliation{Center for High Energy Physics: Kyungpook National University, Daegu 702-701, Korea; Seoul National University, Seoul 151-742, Korea; Sungkyunkwan University, Suwon 440-746, Korea; Korea Institute of Science and Technology Information, Daejeon 305-806, Korea; Chonnam National University, Gwangju 500-757, Korea; Chonbuk National University, Jeonju 561-756, Korea; Ewha Womans University, Seoul, 120-750, Korea}
\affiliation{Ernest Orlando Lawrence Berkeley National Laboratory, Berkeley, California 94720, USA}
\affiliation{University of Liverpool, Liverpool L69 7ZE, United Kingdom}
\affiliation{University College London, London WC1E 6BT, United Kingdom}
\affiliation{Centro de Investigaciones Energeticas Medioambientales y Tecnologicas, E-28040 Madrid, Spain}
\affiliation{Massachusetts Institute of Technology, Cambridge, Massachusetts 02139, USA}
\affiliation{University of Michigan, Ann Arbor, Michigan 48109, USA}
\affiliation{Michigan State University, East Lansing, Michigan 48824, USA}
\affiliation{Institution for Theoretical and Experimental Physics, ITEP, Moscow 117259, Russia}
\affiliation{University of New Mexico, Albuquerque, New Mexico 87131, USA}
\affiliation{The Ohio State University, Columbus, Ohio 43210, USA}
\affiliation{Okayama University, Okayama 700-8530, Japan}
\affiliation{Osaka City University, Osaka 558-8585, Japan}
\affiliation{University of Oxford, Oxford OX1 3RH, United Kingdom}
\affiliation{Istituto Nazionale di Fisica Nucleare, Sezione di Padova, \ensuremath{^{ll}}University of Padova, I-35131 Padova, Italy}
\affiliation{University of Pennsylvania, Philadelphia, Pennsylvania 19104, USA}
\affiliation{Istituto Nazionale di Fisica Nucleare Pisa, \ensuremath{^{mm}}University of Pisa, \ensuremath{^{nn}}University of Siena, \ensuremath{^{oo}}Scuola Normale Superiore, I-56127 Pisa, Italy, \ensuremath{^{pp}}INFN Pavia, I-27100 Pavia, Italy, \ensuremath{^{qq}}University of Pavia, I-27100 Pavia, Italy}
\affiliation{University of Pittsburgh, Pittsburgh, Pennsylvania 15260, USA}
\affiliation{Purdue University, West Lafayette, Indiana 47907, USA}
\affiliation{University of Rochester, Rochester, New York 14627, USA}
\affiliation{The Rockefeller University, New York, New York 10065, USA}
\affiliation{Istituto Nazionale di Fisica Nucleare, Sezione di Roma 1, \ensuremath{^{rr}}Sapienza Universit\`{a} di Roma, I-00185 Roma, Italy}
\affiliation{Mitchell Institute for Fundamental Physics and Astronomy, Texas A\&M University, College Station, Texas 77843, USA}
\affiliation{Istituto Nazionale di Fisica Nucleare Trieste, \ensuremath{^{ss}}Gruppo Collegato di Udine, \ensuremath{^{tt}}University of Udine, I-33100 Udine, Italy, \ensuremath{^{uu}}University of Trieste, I-34127 Trieste, Italy}
\affiliation{University of Tsukuba, Tsukuba, Ibaraki 305, Japan}
\affiliation{Tufts University, Medford, Massachusetts 02155, USA}
\affiliation{Waseda University, Tokyo 169, Japan}
\affiliation{Wayne State University, Detroit, Michigan 48201, USA}
\affiliation{University of Wisconsin-Madison, Madison, Wisconsin 53706, USA}
\affiliation{Yale University, New Haven, Connecticut 06520, USA}

\author{T.~Aaltonen}
\affiliation{Division of High Energy Physics, Department of Physics, University of Helsinki, FIN-00014, Helsinki, Finland; Helsinki Institute of Physics, FIN-00014, Helsinki, Finland}
\author{S.~Amerio\ensuremath{^{ll}}}
\affiliation{Istituto Nazionale di Fisica Nucleare, Sezione di Padova, \ensuremath{^{ll}}University of Padova, I-35131 Padova, Italy}
\author{D.~Amidei}
\affiliation{University of Michigan, Ann Arbor, Michigan 48109, USA}
\author{A.~Anastassov\ensuremath{^{w}}}
\affiliation{Fermi National Accelerator Laboratory, Batavia, Illinois 60510, USA}
\author{A.~Annovi}
\affiliation{Laboratori Nazionali di Frascati, Istituto Nazionale di Fisica Nucleare, I-00044 Frascati, Italy}
\author{J.~Antos}
\affiliation{Comenius University, 842 48 Bratislava, Slovakia; Institute of Experimental Physics, 040 01 Kosice, Slovakia}
\author{G.~Apollinari}
\affiliation{Fermi National Accelerator Laboratory, Batavia, Illinois 60510, USA}
\author{J.A.~Appel}
\affiliation{Fermi National Accelerator Laboratory, Batavia, Illinois 60510, USA}
\author{T.~Arisawa}
\affiliation{Waseda University, Tokyo 169, Japan}
\author{A.~Artikov}
\affiliation{Joint Institute for Nuclear Research, RU-141980 Dubna, Russia}
% ...

\collaboration{CDF Collaboration}
\altaffiliation[With visitors from]{
\ensuremath{^{a}}University of British Columbia, Vancouver, BC V6T 1Z1, Canada,
\ensuremath{^{b}}Istituto Nazionale di Fisica Nucleare, Sezione di Cagliari, 09042 Monserrato (Cagliari), Italy,
\ensuremath{^{c}}University of California Irvine, Irvine, CA 92697, USA,
\ensuremath{^{d}}Institute of Physics, Academy of Sciences of the Czech Republic, 182~21, Czech Republic,
\ensuremath{^{e}}CERN, CH-1211 Geneva, Switzerland,
\ensuremath{^{f}}Cornell University, Ithaca, NY 14853, USA,
\ensuremath{^{g}}University of Cyprus, Nicosia CY-1678, Cyprus,
\ensuremath{^{h}}Office of Science, U.S. Department of Energy, Washington, DC 20585, USA,
\ensuremath{^{i}}University College Dublin, Dublin 4, Ireland,
\ensuremath{^{j}}ETH, 8092 Z\"{u}rich, Switzerland,
\ensuremath{^{k}}University of Fukui, Fukui City, Fukui Prefecture, Japan 910-0017,
\ensuremath{^{l}}Universidad Iberoamericana, Lomas de Santa Fe, M\'{e}xico, C.P. 01219, Distrito Federal,
\ensuremath{^{m}}University of Iowa, Iowa City, IA 52242, USA,
\ensuremath{^{n}}Kinki University, Higashi-Osaka City, Japan 577-8502,
\ensuremath{^{o}}Kansas State University, Manhattan, KS 66506, USA,
\ensuremath{^{p}}Brookhaven National Laboratory, Upton, NY 11973, USA,
\ensuremath{^{q}}Istituto Nazionale di Fisica Nucleare, Sezione di Lecce, Via Arnesano, I-73100 Lecce, Italy,
\ensuremath{^{r}}Queen Mary, University of London, London, E1 4NS, United Kingdom,
\ensuremath{^{s}}University of Melbourne, Victoria 3010, Australia,
\ensuremath{^{t}}Muons, Inc., Batavia, IL 60510, USA,
\ensuremath{^{u}}Nagasaki Institute of Applied Science, Nagasaki 851-0193, Japan,
\ensuremath{^{v}}National Research Nuclear University, Moscow 115409, Russia,
\ensuremath{^{w}}Northwestern University, Evanston, IL 60208, USA,
\ensuremath{^{x}}University of Notre Dame, Notre Dame, IN 46556, USA,
\ensuremath{^{y}}Universidad de Oviedo, E-33007 Oviedo, Spain,
\ensuremath{^{z}}CNRS-IN2P3, Paris, F-75205 France,
\ensuremath{^{aa}}Universidad Tecnica Federico Santa Maria, 110v Valparaiso, Chile,
\ensuremath{^{bb}}Sejong University, Seoul 143-747, Korea,
\ensuremath{^{cc}}The University of Jordan, Amman 11942, Jordan,
\ensuremath{^{dd}}Universite catholique de Louvain, 1348 Louvain-La-Neuve, Belgium,
\ensuremath{^{ee}}University of Z\"{u}rich, 8006 Z\"{u}rich, Switzerland,
\ensuremath{^{ff}}Massachusetts General Hospital, Boston, MA 02114 USA,
\ensuremath{^{gg}}Harvard Medical School, Boston, MA 02114 USA,
\ensuremath{^{hh}}Hampton University, Hampton, VA 23668, USA,
\ensuremath{^{ii}}Los Alamos National Laboratory, Los Alamos, NM 87544, USA,
\ensuremath{^{jj}}Universit\`{a} degli Studi di Napoli Federico II, I-80138 Napoli, Italy
}
\noaffiliation

%% file: cdf_isr.bbl
\begin{thebibliography}{}
\bibitem{met}{
  The calorimeters have a tower-geometry segmentation with towers pointing to the center of the detector. Transverse energy imbalance $\met$ is the magnitude of $-\sum_i E_T^i\hat{n}_i$, where the sum is over calorimeter towers, $\hat{n}_i$ is the unit vector normal to beam direction and pointing to a given tower, and $E_T^i$ is the transverse energy measured in that tower.
}
\end{thebibliography}
